\documentclass[11pt]{article}

\usepackage[preprint]{acl}

\usepackage{times}
\usepackage{latexsym}
\usepackage[T1]{fontenc}
\usepackage[utf8]{inputenc}
\usepackage{microtype}
\usepackage{inconsolata}

\usepackage{amsmath}
\usepackage{amssymb}
\usepackage{amsfonts}
\usepackage{mathtools}
\usepackage{cases}
\usepackage{graphicx}
\usepackage{booktabs}
\usepackage{multirow}
\usepackage{makecell}
\usepackage{array}
\usepackage{xcolor}
\usepackage{url}
\usepackage{enumitem}
\usepackage{listings}
\usepackage{algorithm}
\usepackage{algorithmic}
\usepackage[capitalize]{cleveref}

\graphicspath{{Figure/}}

\crefname{section}{Sec.}{Secs.}
\Crefname{section}{Section}{Sections}
\crefname{table}{Tab.}{Tabs.}
\Crefname{table}{Table}{Tables}
\crefname{figure}{Fig.}{Figs.}
\Crefname{figure}{Figure}{Figures}

\definecolor{PromptBlue}{HTML}{2F5F9F}
\definecolor{PromptBlueBg}{HTML}{F3F7FC}
\definecolor{PromptBlueBorder}{HTML}{8BAED6}
\definecolor{PromptGreen}{HTML}{2E6B57}
\definecolor{PromptGreenBg}{HTML}{F3FAF7}
\definecolor{PromptGreenBorder}{HTML}{8BC2AE}

\lstdefinestyle{promptcode}{
  basicstyle=\ttfamily\tiny,
  keywordstyle=\ttfamily\bfseries,
  stringstyle=\ttfamily,
  commentstyle=\ttfamily,
  breaklines=true,
  breakatwhitespace=false,
  columns=fullflexible,
  keepspaces=true,
  frame=single,
  framerule=0.2pt,
  rulecolor=\color{black!25},
  xleftmargin=0.5em,
  xrightmargin=0.5em,
  aboveskip=0.75em,
  belowskip=0.75em
}

\lstdefinestyle{promptscreening}{
  style=promptcode,
  backgroundcolor=\color{PromptBlueBg},
  rulecolor=\color{PromptBlueBorder}
}

\lstdefinestyle{promptcleaning}{
  style=promptcode,
  backgroundcolor=\color{PromptGreenBg},
  rulecolor=\color{PromptGreenBorder}
}

\newcommand{\prompttitlebar}[2]{%
  \noindent\colorbox{#1}{%
    \parbox{\dimexpr\textwidth-2\fboxsep\relax}{%
      \strut\color{white}\textbf{\small #2}%
    }%
  }%
  \vspace{-0.4em}%
}

\title{Auditing and Mitigating Privacy Leakage in Cloud-Edge Collaborative Decoding}

\author{
  Kejia Zhang\textsuperscript{1} \quad
  Tianyuan Zou\textsuperscript{2} \quad
  Zixuan Gu\textsuperscript{3} \quad
  Yang Liu\textsuperscript{1}\thanks{Corresponding author: \href{mailto:yang-veronica.liu@polyu.edu.hk}{yang-veronica.liu@polyu.edu.hk}.} \\
  \textsuperscript{1}The Hong Kong Polytechnic University \\
  \textsuperscript{2}Institute for AI Industry Research, Tsinghua University \\
  \textsuperscript{3}School of Software, Tsinghua University
}

\begin{document}
\maketitle

\begin{abstract}
  Applications such as personalized assistance and proprietary document analysis require large language models (LLMs) to generate outputs from private data.
  Yet powerful LLMs typically cannot be deployed on the resource-constrained devices where private data resides, and uploading private data to cloud-hosted LLMs exposes sensitive information.
  Recent work addresses this tension with a \textit{cloud-edge collaborative decoding} paradigm, where private data are kept on the edge with a small language model (SLM) producing next-token distributions, which are fused with predictions from a cloud LLM operating solely on public data.
  In this paper, we systematically analyze the privacy risks of such a paradigm with a novel evaluation framework using constructed QA datasets,
  which show that such collaboration can expose substantial private-context information.
  To address such privacy leakage, we propose \textsc{CoVeil}, a defense mechanism which dynamically optimizes transmitted signals to suppress leakage during decoding time while preserving the collaborative quality.
  Extensive evaluations demonstrate that \textsc{CoVeil} consistently improves the privacy-utility trade-off over existing baselines by reducing data leakage by up to 87.2\%, with minimal accuracy loss.
\end{abstract}

\section{Introduction}

Large language models (LLMs) have emerged as powerful tools for reasoning over private data, with applications spanning clinical decision support~\citep{hager2024evaluation}, personal assistance~\citep{salemi2025lamp}, and proprietary document analysis~\citep{ma2024mmlongbench}.
However, deploying LLMs raises a core conflict between capability and privacy.
Powerful LLMs are computationally expensive to run on edge devices that private data often resides~\citep{lu2025demystifying,fu2024amoeballm},
while uploading private data to cloud-hosted LLMs exposes it to external service providers~\citep{tong2025inferdpt}.
Prior work mitigates this tension by deploying small language models (SLMs) through knowledge distillation, model compression~\citep{guminillm,sunsimple,huostquant,feng2026dmoe,tuo2025sparsessm}, or by sanitizing private prompts before cloud inference~\citep{tong2025inferdpt,zhang2025dyntext}.
However, these approaches limit task utility by relying on smaller local models or removing useful context before cloud inference.

To better leverage LLM capability without uploading the private context, an important research direction couples a cloud-hosted LLM with a locally deployed SLM, allowing the two models to collaborate through decoding signals while the cloud LLM operates only on public or non-sensitive data and the SLM retains the private context~\citep{zhang2024cogenesis,zhengciter,she2025token,liu2025towards}.
We refer to this paradigm as \textit{cloud-edge collaborative decoding}.
At each autoregressive step, the edge SLM and the cloud LLM each produce a next-token distribution, and the two distributions are fused to generate the next token.

\begin{figure*}[t]
    \centering
    \includegraphics[width=0.75\textwidth]{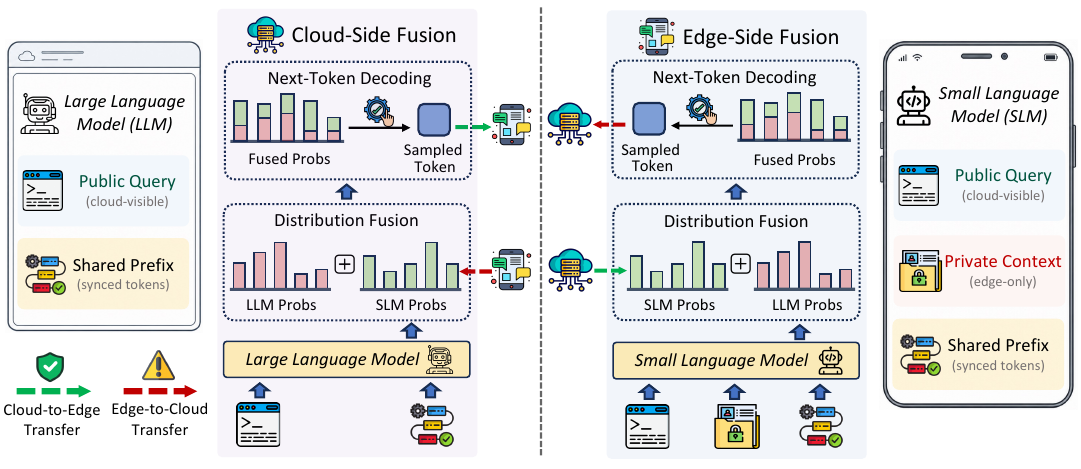}
    \vspace{-0.8em}
        \caption{
        Cloud-edge collaborative decoding, where private context remains on the edge under two fusion modes.
        }
    \label{fig:intro_teaser_reconstruct}
    \vspace{-0.5em}
\end{figure*}

Depending on where the two next-token distributions are fused, this paradigm can be categorized into \textbf{cloud-side fusion}, where the edge SLM uploads its next-token probabilities (probability-level signal) to the cloud~\citep{zhang2024cogenesis}, and \textbf{edge-side fusion}, where the edge locally fuses the two distributions and synchronizes the sampled token (token-level signal) to the cloud~\citep{zhengciter,she2025token}.
Although prior work has shown that next-token probabilities can leak information about input data~\citep{morris2024language,finlaysonlogits,nazir2025better}, a systematic analysis of the distinct privacy risks inherent to both probability-level and token-level signals remains largely absent.

Motivated by this, we systematically analyze whether the cloud-observed decoding signals during collaborative decoding leak the private edge-side context.
We introduce \textsc{DecodeLeak}, a privacy evaluation framework for cloud-edge collaborative decoding, and construct two QA benchmarks from medical and commonsense QA sources~\citep{jin2021disease,huang2019cosmos};
each instance pairs a public query with an edge-side private context.
Each data sample also includes manually annotated private evidence, defined as a set of critical sensitive text spans in the private context needed for answering
the public query, and are used to quantify the level of leakage.
For example, for a medication-selection query, private context spans such as ``respiratory depression'', ``pupillary constriction'' are annotated as private evidence because they are needed to select the correct medication.
\textsc{DecodeLeak}
evaluates leakage in two sequential steps:
\textit{private evidence recall}, which measures the fraction of annotated private evidence exposed by the top-$K$ token IDs from the per-step next-token distributions;
then \textit{private context inversion} which uses the corresponding per-step top-$K$ exposure sequence as the attack input for reconstructing the private context.
We find that cloud-observed decoding signals can reveal substantial private evidence using our evaluation framework.

These findings motivate us to develop privacy-preserving collaborative decoding mechanisms.
Existing defenses typically apply fixed global controls to uploaded distributions~\citep{zhang2024cogenesis,thareja2026dp}, or use privacy-aware rules to choose the synchronized token~\citep{ghoukasian2026locally,vinod2026invisibleink}.
Such controls may suppress utility-critical signals or retain leaky low-utility ones.
We therefore propose \textsc{CoVeil}, which optimizes each cloud-observed signal under a utility-privacy trade-off at decoding time, enabling adaptive selection of upload probabilities in cloud-side fusion, or a synchronized token in edge-side fusion.
Experiments show that per-step utility-aware signal selection reduces private evidence leakage while preserving task utility.

In summary, this work makes three contributions.
First, we identify and analyze privacy risks of cloud-edge collaborative decoding, showing that uploaded SLM probabilities and synchronized tokens can expose private evidence and enable private context inversion.
Second, we present \textsc{DecodeLeak}, a privacy evaluation framework, and construct two QA benchmarks with public queries, private edge-side contexts, and private-evidence annotations.
Third, we propose \textsc{CoVeil}, a post-hoc privacy-aware collaborative method that optimizes cloud-observed decoding signals at decoding time without incurring additional model training, and show that it significantly improves the privacy-utility trade-off over existing baselines.

\section{Problem Setup and Threat Model}

\subsection{Cloud-Edge Collaborative Decoding}
\label{sec:setting}

In cloud-edge collaborative decoding, a cloud-hosted LLM and an
edge-side SLM jointly perform next token prediction while the private context $x_{\mathrm{priv}}$
remains solely on the edge and the public query $x_{\mathrm{pub}}$ is shared by both models~\citep{zhang2024cogenesis,tian2026floe,she2025token}, as shown in \Cref{fig:intro_teaser_reconstruct}.
Let $\mathcal{V}$ denote the vocabulary shared between SLM and LLM.
At each decoding step $t$, the cloud LLM and the edge SLM produce next-token
probability distributions over $\mathcal{V}$:
\begin{numcases}{}
  p_L^{(t)}(\cdot)
  =
  f_{\theta_L}\!\left(\cdot \mid y_{<t}, x_{\mathrm{pub}}\right),\\
  p_S^{(t)}(\cdot)
  =
  f_{\theta_S}\!\left(\cdot \mid y_{<t}, x_{\mathrm{pub}}, x_{\mathrm{priv}}\right),
\end{numcases}
where $f_{\theta_L}$ and $f_{\theta_S}$ denote the cloud LLM and edge SLM parameterized by $\theta_L$ and $\theta_S$, respectively;
$p_L^{(t)}(\cdot)$ and $p_S^{(t)}(\cdot)$ denote their next-token probability distributions over the vocabulary $\mathcal{V}$ at step $t$;
$y_{<t}$ denotes the shared autoregressive prefix.
The fused next-token distribution is typically obtained by a weighted
interpolation~\citep{zhang2024cogenesis}:
\vspace{-0.5em}
\begin{equation}
  p_{\mathrm{fuse}}^{(t)}
  =
  \alpha \cdot p_L^{(t)}
  +
  (1-\alpha) \cdot p_S^{(t)} ,
\end{equation}
where $\alpha \in [0,1]$ is the fusion weight.
The next token is then sampled as $y_t^{\mathrm{fuse}} \sim p_{\mathrm{fuse}}^{(t)}(\cdot)$.

\subsection{Fusion Modes}
\label{sec:fusion_carriers}

Let $o_L^{(t)}$ denote the cloud-observed decoding signal at step $t$.
The fusion mode determines $o_L^{(t)}$ as:
\vspace{-0.5em}
\begin{equation}
  o_L^{(t)}
  =
  \begin{cases}
    p_S^{(t)}, & \text{cloud-side fusion},\\
    y_t^{\mathrm{fuse}}, & \text{edge-side fusion}.
  \end{cases}
\end{equation}

\paragraph{Cloud-Side Fusion (Probability-Level Signal).}
The edge uploads the SLM next-token probabilities $p_S^{(t)}$ to the cloud,
which fuses them with the LLM distribution $p_L^{(t)}$ and samples
$y_t^{\mathrm{fuse}}$ from the fused distribution~\citep{zhang2024cogenesis,tian2026floe}.

\paragraph{Edge-Side Fusion (Token-Level Signal).}
The cloud sends $p_L^{(t)}$ to the edge, and the edge fuses it
with $p_S^{(t)}$ and samples $y_t^{\mathrm{fuse}}$~\citep{zhengciter,she2025token}.
The sampled token is synchronized to the cloud and appended to the shared
autoregressive prefix, thereby conditioning the cloud LLM's subsequent
next-token distribution.

\subsection{Threat Model}
\label{sec:threat_model}

We assume that the cloud LLM is honest-but-curious: it follows the decoding protocol but attempts to infer the private context from its cloud-side observations~\citep{qu2025prompt,ding2024patrol}. After $T$ decoding steps, the cloud has access to the public query $x_{\mathrm{pub}}$, its own parameters $\theta_L$, the generated prefix $y_{\leq T}^{\mathrm{fuse}}$, and the cloud-observed decoding signal sequence $\{o_L^{(t)}\}_{t=1}^{T}$.

This induces two leakage channels depending on the fusion mode: in cloud-side fusion, \textbf{probability-level leakage} arises because the cloud observes the uploaded SLM next-token probabilities $p_S^{(t)}$; in edge-side fusion, \textbf{token-level leakage} arises because the cloud observes the synchronized token $y_t^{\mathrm{fuse}}$ in the shared prefix and can compute its own prefix-conditioned distribution $p_L^{(t)}(\cdot \mid y_{<t}^{\mathrm{fuse}}, x_{\mathrm{pub}})$.
In both cases, the cloud attempts to infer private information $x_{\mathrm{priv}}$ from these observations.

\section{\textsc{DecodeLeak}}
\label{sec:signal_leakage}

\begin{figure*}[t]
\centering
\includegraphics[width=0.92\textwidth]{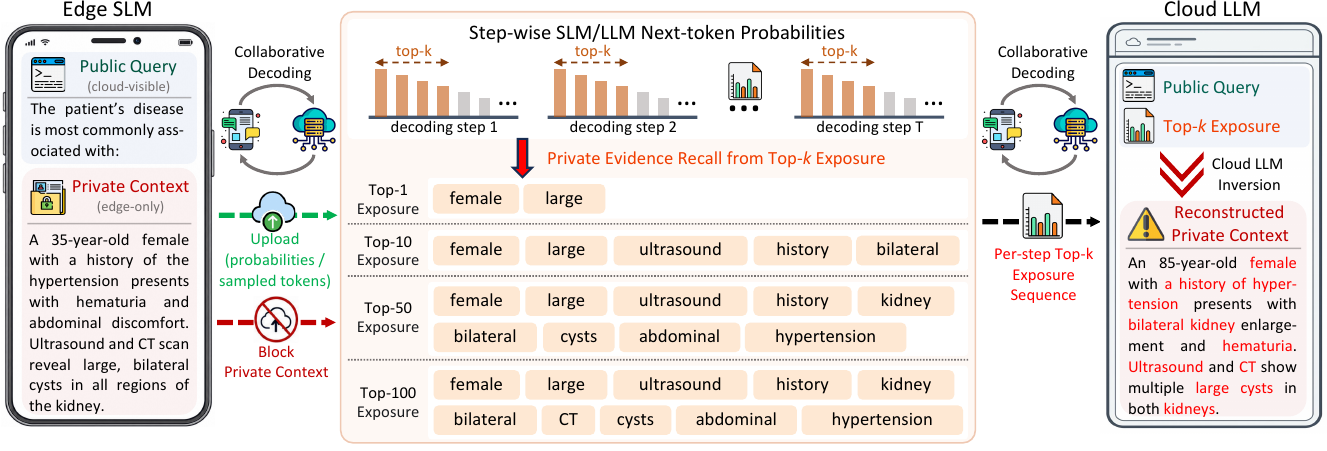}
\vspace{-0.7em}
\caption{\textsc{DecodeLeak} overview: cloud-observed decoding signals are evaluated in two sequential stages: private evidence recall from top-$K$ token IDs, followed by private context inversion from the per-step exposure sequence.}
\vspace{-0.7em}
\label{fig:decodeleak_attack_schematic}
\end{figure*}

\textsc{DecodeLeak} first defines annotated private evidence as the evaluation target, then evaluates leakage in two sequential steps: private evidence recall over cloud-observed top-$K$ token sets, followed by private context inversion from the corresponding per-step exposure sequence.

\subsection{Private Evidence and Benchmark}
\label{sec:private_evidence}
Prior collaborative decoding studies focus on task performance and efficiency without private-evidence annotations~\citep{zhang2024cogenesis,tian2026floe}, while privacy and inversion studies typically focus on single-model inference rather than the cloud-edge setting with a public/private input split~\citep{qu2025prompt,flemings2025estimating}. We therefore construct benchmarks that explicitly separate public queries, private edge-side contexts, and annotated private evidence.

We define the evaluation instance for \textsc{DecodeLeak} as a tuple $(x_{\mathrm{pub}}, x_{\mathrm{priv}}, \mathcal{E}_{\mathrm{priv}})$, where $x_{\mathrm{pub}}$ is the public query available to the cloud, $x_{\mathrm{priv}}$ is the private context retained on the edge, and $\mathcal{E}_{\mathrm{priv}}=\{P_1,\ldots,P_M\}$ is the private-evidence set.
Each span $P_m \subseteq x_{\mathrm{priv}}$ contains private text that is needed to answer $x_{\mathrm{pub}}$ but absent from $x_{\mathrm{pub}}$.

We instantiate this definition with two multiple-choice QA benchmarks: \textsc{MedPriv} (1{,}000 medical QA instances~\citep{jin2021disease}) and \textsc{CommPriv} (1{,}000 commonsense QA instances~\citep{huang2019cosmos}).
In each instance, $x_{\mathrm{pub}}$ contains the question and answer options $\mathcal{A}$, while $x_{\mathrm{priv}}$ is a supporting passage with correct option $a^\star$.
We call $\mathcal{E}_{\mathrm{priv}}$ answer-critical if $x_{\mathrm{pub}}$ is insufficient to select $a^\star$ from $\mathcal{A}$, but $(x_{\mathrm{pub}},\, \mathcal{E}_{\mathrm{priv}})$ is sufficient.

\noindent \paragraph{Private evidence annotation.}
The private-evidence set $\mathcal{E}_{\mathrm{priv}}$ is obtained in two stages.
First, we use GPT-5.4 to propose candidate instances where the correct option $a^\star$ can be determined from $(x_{\mathrm{pub}}, x_{\mathrm{priv}})$ but not from $x_{\mathrm{pub}}$ alone.
Second, we extract minimal private spans from $x_{\mathrm{priv}}$ that are sufficient to identify $a^\star$; both stages are manually verified.
For example, in a diagnosis query, private-context spans such as ``sensitive to criticism'' and ``socially inhibited'' are annotated as private evidence because they support the diagnosis.
Beyond this QA instantiation, \textsc{DecodeLeak} can be applied to datasets that can be split into a public query and a private context with annotated private evidence.
Detailed construction and annotation procedures are provided in \Cref{app:benchmark_construction,app:benchmark_prompts}.

\subsection{Private Evidence Recall}
\label{sec:coverage}
Private evidence recall is the first step of \textsc{DecodeLeak}; it quantifies leakage as the fraction of annotated private evidence exposed by cloud-observed top-$K$ token sets extracted from per-step next-token distributions.
At step $t$, the signal exposure set is:
\vspace{-0.5em}
\begin{equation}
\vspace{-0em}
\small
E_t^K =
\begin{cases}
\operatorname{TopK}(p_S^{(t)}), & \text{probability-level,}\\
\operatorname{TopK}\!\left(p_L^{(t)}(\cdot \mid y_{<t}^{\mathrm{fuse}}, x_{\mathrm{pub}})\right), & \text{token-level,}
\end{cases}
\end{equation}
where $\operatorname{TopK}(\cdot)$ returns the $K$ token IDs with the highest probability.
$E^K = \bigcup_{t=1}^{T} E_t^K$ aggregates the exposed token IDs across $T$ steps, while the sequence $\{E_t^K\}_{t=1}^{T}$ is used for private context inversion in \Cref{sec:reconstruction}.
A higher recall therefore indicates that more annotated private evidence is directly present in the decoding signals observed by the cloud, suggesting more severe privacy leakage.

We report recall-style metrics between $E^K$ and $\mathcal{E}_{\mathrm{priv}}$ from token-, word-, and span-level views~\citep{thareja2026dp,vinod2026invisibleink}.
Let $\mathcal{T}(\cdot)$ tokenize a text segment or a set of spans into its set of content token IDs under $\mathcal{V}$, excluding stopwords and punctuation.
Let $\mathcal{W}(\cdot)$ return the content word set, decoded from token IDs when applied to $E^K$ and tokenized from text otherwise.

\noindent \textbf{Token Evidence Recall@K} quantifies overlap at the token-ID level:
\vspace{-0.5em}
\begin{equation}
\mathrm{Token\text{-}ER@}K =
\frac{|\mathcal{T}(\mathcal{E}_{\mathrm{priv}}) \cap E^K|}{|\mathcal{T}(\mathcal{E}_{\mathrm{priv}})|}.
\end{equation}

\noindent \textbf{ROUGE-1 Evidence Recall@K} captures lexical overlap between the decoded top-$K$ token set and the private-evidence text at the word level:
\begin{equation}
\mathrm{ROUGE1\text{-}ER@}K =
\frac{|\mathcal{W}(\mathcal{E}_{\mathrm{priv}})\cap\mathcal{W}(E^K)|}{|\mathcal{W}(\mathcal{E}_{\mathrm{priv}})|}.
\end{equation}

\noindent \textbf{Span Evidence Recall@K} reports recall averaged over the private-evidence spans:
\begin{equation}
\mathrm{Span\text{-}ER@}K =
\frac{1}{M}
\sum_{m=1}^{M}
\frac{|\mathcal{T}(P_m) \cap E^K|}{|\mathcal{T}(P_m)|}.
\end{equation}

\noindent \textbf{AUC@$K$} summarizes Token-ER across cutoffs up to $K$ to reduce sensitivity to a specific cutoff:
\begin{equation}
\mathrm{AUC@}K = \frac{1}{K}\sum_{k=1}^{K}\mathrm{Token\text{-}ER@}k.
\end{equation}

\begin{figure*}[t]
\centering
\begin{minipage}[t]{0.49\textwidth}
    \centering
    \includegraphics[width=\linewidth]{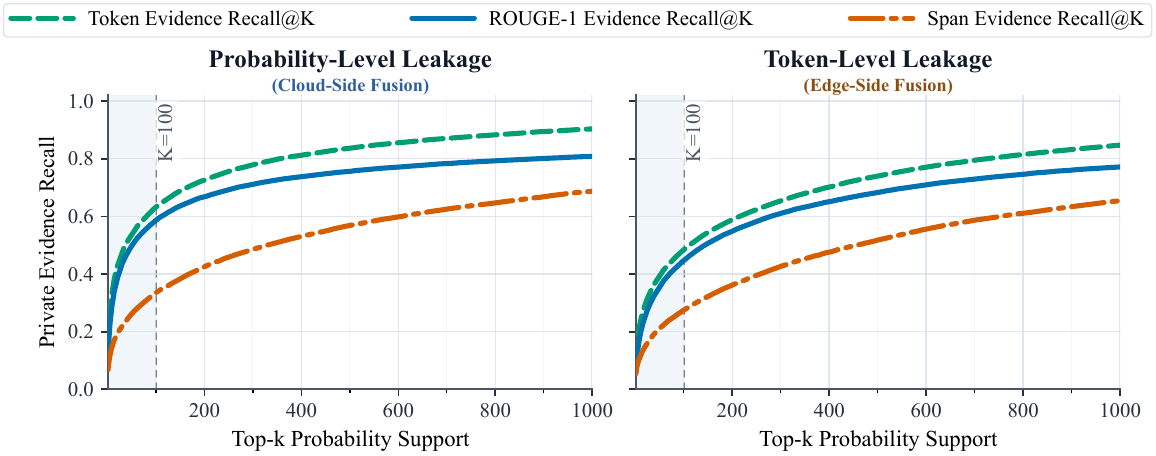}
    \vspace{-0.6em}
    {\small (a) \textsc{MedPriv}}
\end{minipage}
\hfill
\begin{minipage}[t]{0.49\textwidth}
    \centering
    \includegraphics[width=\linewidth]{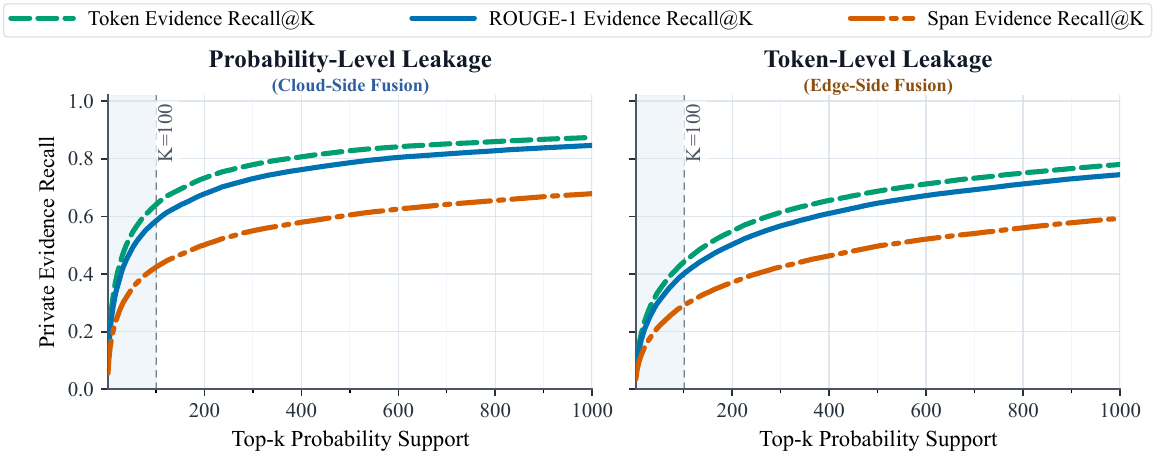}
    \vspace{-0.6em}
    {\small (b) \textsc{CommPriv}}
\end{minipage}
\caption{Private evidence recall across leakage channels in Qwen2.5-72B/1.5B collaboration using 200 examples.}
\label{fig:trace_coverage_k1000}
\end{figure*}

As $K$ increases, private evidence recall consistently rises across both fusion modes and datasets (\Cref{fig:trace_coverage_k1000}).
At $K=100$ (fewer than 0.1\% of the Qwen vocabulary), Token-ER reaches approximately 60\% of private-evidence tokens when the cloud observes uploaded SLM probabilities in cloud-side fusion, and 50\% when it observes synchronized tokens in edge-side fusion.

\subsection{Private Context Inversion}
\label{sec:reconstruction}

Private context inversion measures whether the cloud can reconstruct $x_{\mathrm{priv}}$ from the signal exposure sequence $\{E_t^K\}_{t=1}^{T}$ and $x_{\mathrm{pub}}$, as illustrated in the right panel of \Cref{fig:decodeleak_attack_schematic}.

Given $x_{\mathrm{pub}}$ and the signal exposure sequence $\{E_t^K\}_{t=1}^{T}$, the cloud constructs an inversion prompt from $(x_{\mathrm{pub}}, \{E_t^K\}_{t=1}^{T})$\footnote{The inversion prompt template is shown in \Cref{fig:inversion_prompt}.} and uses the cloud LLM to reconstruct the private context $\hat{x}_{\mathrm{priv}}$:
\begin{equation}
  \hat{x}_{\mathrm{priv}} = f_{\theta_L}\!\left(\cdot \mid \pi\!\left(x_{\mathrm{pub}},\ \{E_t^K\}_{t=1}^{T}\right)\right)
\end{equation}

Inversion quality is quantified by comparing $\hat{x}_{\mathrm{priv}}$ with the ground-truth $x_{\mathrm{priv}}$.

\noindent \textbf{ROUGE-1 Recall} measures the fraction of private-context unigrams recovered by the inversion.

\noindent \textbf{ROUGE-1 F1} measures balanced unigram overlap between the reconstructed and full private contexts~\citep{lin2004rouge}.

As reported in \Cref{tab:decodeleak_results}, the exposure sequence enables substantial private context inversion, with ROUGE-1 Recall of 41.1\% under cloud-side fusion and 39.2\% under edge-side fusion.
Together with the private evidence recall results above, this shows that cloud-observed decoding signals expose private information at both the token-exposure and text-reconstruction levels, motivating privacy-aware optimization.

\section{\textsc{CoVeil}: Privacy-Aware Fusion}

To reduce privacy leakage from uploaded distributions and synchronized tokens, we propose \textsc{CoVeil}, a privacy-aware defense that optimizes cloud-observed decoding signals at decoding time while preserving decoding utility.
\textsc{CoVeil} runs on the edge in both fusion modes, using the cloud-provided $p_L^{(t)}$ together with the edge-local $p_S^{(t)}$.
In cloud-side fusion, the edge uploads only the selected probability--ID pairs $\{(i,p_S^{(t)}(i)):i\in\mathcal{S}_t\}$; in edge-side fusion, it performs fusion locally and synchronizes only the selected token $i_t^\star$.
Thus, the full SLM distribution $p_S^{(t)}$ remains on the edge in both modes.

\subsection{Stepwise Privacy-Utility Optimization}

At each decoding step, \textsc{CoVeil} optimizes the candidate cloud-observed decoding signal $o_t$ by directly balancing utility against privacy cost:
\begin{equation}
  \min_{o_t}
  \mathcal{J}_t(o_t)
  =
  -\mathcal{U}_t(o_t)
  +
  \lambda \cdot \mathcal{P}_t(o_t),
\end{equation}
where $o_t$ denotes the candidate realization of the cloud-observed signal $o_L^{(t)}$ optimized by \textsc{CoVeil};
$\mathcal{U}_t$ and $\mathcal{P}_t$ measure decoding utility and privacy cost, respectively; and $\lambda > 0$ denotes the trade-off parameter.
The concrete form of $o_t$, $\mathcal{U}_t$, and $\mathcal{P}_t$ depends on the fusion mode, as detailed below.

\subsection{Cloud-Side Fusion Defense}

Cloud-side fusion exposes the SLM next-token probabilities to the cloud~\citep{morris2024language,finlaysonlogits}.
\textsc{CoVeil} reduces this leakage through adaptive sparse upload of selected SLM next-token probabilities.
For efficiency, we restrict upload decisions to $\mathcal{C}_t=\operatorname{TopK}(p_S^{(t)})$, since low-probability positions add many optimization variables but contribute little to the fused distribution.
A sensitivity study is provided in \Cref{app:qwen_results}.

Specifically, we introduce a selective upload mask $m_t$ to determine which token positions in the SLM next-token probabilities are uploaded
:
\begin{equation}
m_t(i)\in
\begin{cases}
[0,1], & i\in\mathcal{C}_t,\\
\{0\}, & i\notin\mathcal{C}_t.
\end{cases}
\end{equation}
where $[0,1]$ enables gradient computation; the final upload set is binary.
With the upload mask $m_t$, the per-step objective can be written as:
\begin{equation}
  \min_{m_t \in [0,1]^{|\mathcal{C}_t|}}
  \mathcal{J}_t(m_t)
  =
  -\mathcal{U}_t(m_t)
  +
  \lambda \cdot \mathcal{P}_t(m_t).
\end{equation}

\textbf{Utility function.}
The utility term assigns higher value to upload masks that preserve the next-token choice of the unmasked full-fusion distribution by increasing its margin over competing candidates~\citep{zhang2025towards}.
Let $y_t^\star=\arg\max_i p_{\mathrm{fuse}}^{(t)}(i)$ denote the top-ranked token under the unmasked fused distribution.
For each candidate token $j\in\mathcal{C}_t$, $M_t(m_t;j)$ compares the masked fusion score of $y_t^\star$ with that of $j$:
\begin{equation}
  M_t(m_t;j) = s_t(y_t^\star;m_t)-s_t(j;m_t).
\end{equation}
where $s_t(i;m_t)=\alpha \cdot p_L^{(t)}(i)+(1-\alpha)\cdot m_t(i)\cdot p_S^{(t)}(i)$ is the masked fusion score, and $m_t(i)$ controls the SLM probability upload.

The per-step utility is the average margin gain over the zero-upload baseline ($m_t=\mathbf{0}$):
\begin{equation}
  \footnotesize
  \label{eq:cloud_side_utility}
  \mathcal{U}_t(m_t)=
  \frac{1}{|\mathcal{C}_t|}
  \sum_{j\in\mathcal{C}_t}
  \bigl(M_t(m_t;j)-M_t(\mathbf{0};j)\bigr),
\end{equation}
where $M_t(\mathbf{0};j)$ denotes the margin when no SLM token probability is uploaded.
This utility metric requires no labeled data or external supervision and is computed solely from the collaborative decoding distributions at each step.

\textbf{Privacy function.}
\textsc{CoVeil} estimates private-context influence by the discrepancy between edge and cloud token distributions:
\begin{equation}
  \mathcal{P}_t(m_t)
  =
  \sum_{i\in\mathcal{C}_t}
  m_t(i)
  \left|
  p_S^{(t)}(i)-p_L^{(t)}(i)
  \right|.
  \label{eq:cloud_side_privacy}
\end{equation}
The absolute value penalizes the magnitude of the edge-cloud deviation, regardless of its direction.

\textbf{Upload rule.}
To decide which SLM probabilities to upload, \textsc{CoVeil} uses a first-order Taylor expansion of the objective around the zero-upload baseline $m_t=\mathbf{0}$:
\begin{equation}
  \label{eq:upload_taylor}
  \mathcal{J}_t(m_t)
  \approx
  \mathcal{J}_t(\mathbf{0})
  +
  \sum_{i\in\mathcal{C}_t}
  g_t(i)\,m_t(i),
\end{equation}
where $g_t(i) = \left.\frac{\partial \mathcal{J}_t(m_t)}{\partial m_t(i)}\right|_{m_t=\mathbf{0}}$ is the marginal change in $\mathcal{J}_t$ from uploading the SLM probability for token $i$.
Since \textsc{CoVeil} minimizes $\mathcal{J}_t$, $g_t(i)<0$ indicates that uploading token $i$ decreases the objective, whereas $g_t(i)\ge 0$ indicates no improvement under the linear approximation.
Therefore, the uploaded set is:
\vspace{-0.5em}
\begin{equation}
  \mathcal{S}_t
  =
  \{i\in\mathcal{C}_t \mid g_t(i)<0\}.
\end{equation}

\begin{figure*}[t]
\centering
\setlength{\tabcolsep}{0pt}
\begin{tabular}{@{}cccc@{}}
\includegraphics[width=0.249\textwidth]{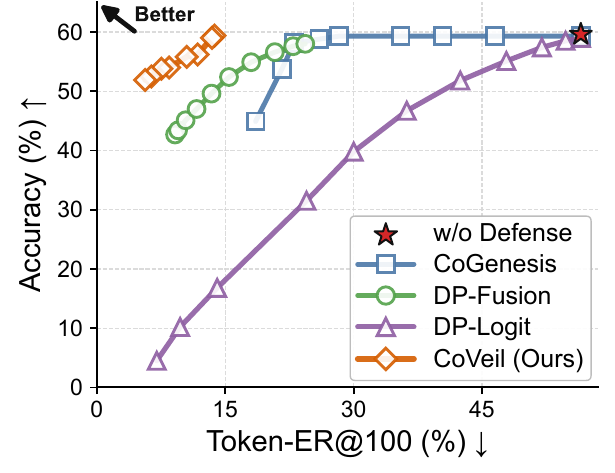} &
\includegraphics[width=0.249\textwidth]{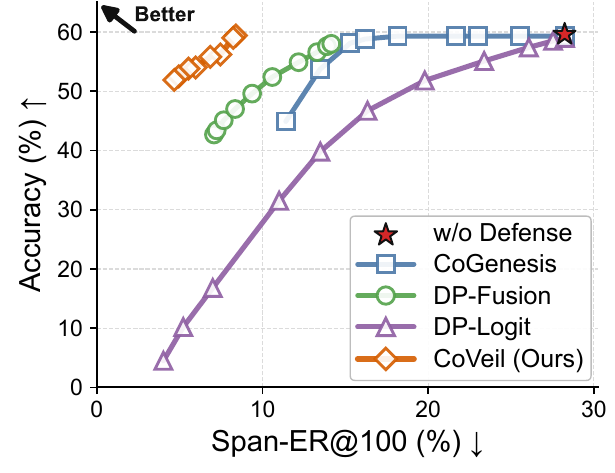} &
\includegraphics[width=0.249\textwidth]{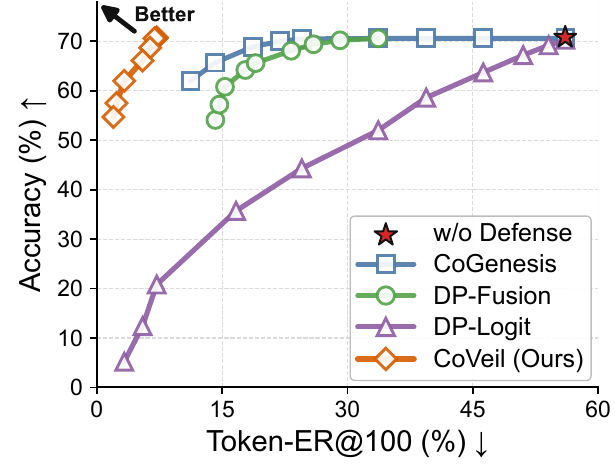} &
\includegraphics[width=0.249\textwidth]{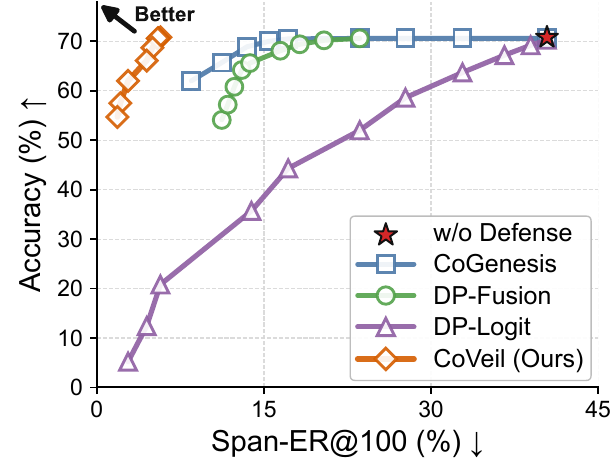} \\[-0.25em]
\multicolumn{2}{c}{\small (a) \textsc{MedPriv}: Cloud-side fusion defense} &
\multicolumn{2}{c}{\small (b) \textsc{CommPriv}: Cloud-side fusion defense} \\[-0.0em]
\includegraphics[width=0.249\textwidth]{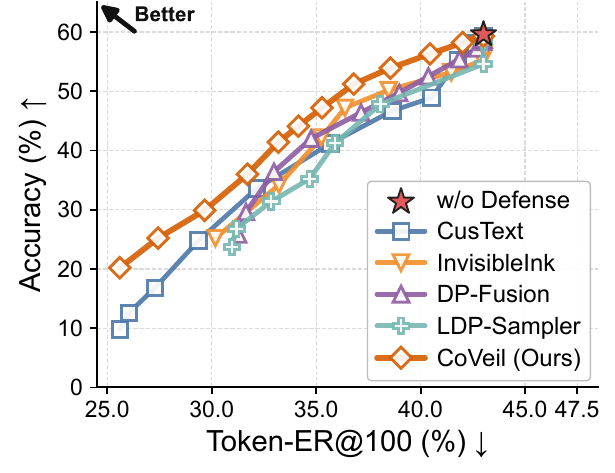} &
\includegraphics[width=0.249\textwidth]{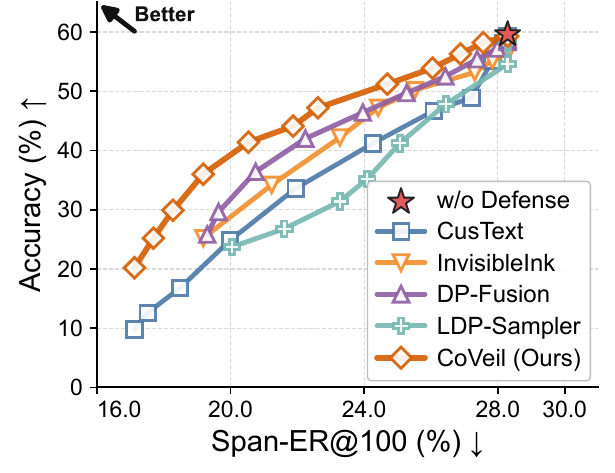} &
\includegraphics[width=0.249\textwidth]{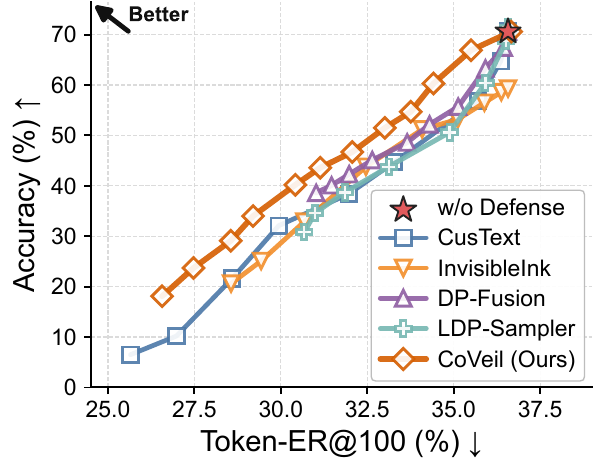} &
\includegraphics[width=0.249\textwidth]{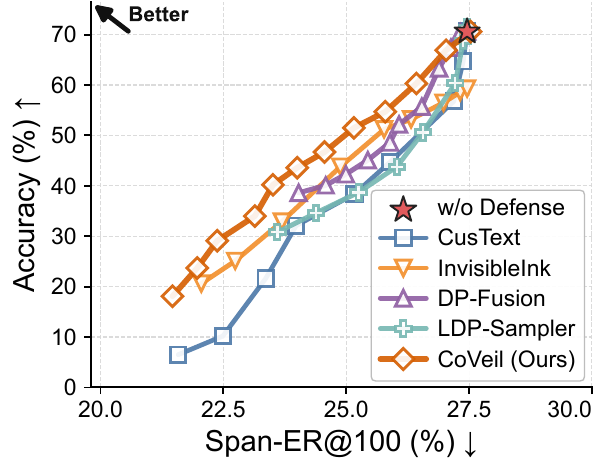} \\[-0.25em]
\multicolumn{2}{c}{\small (c) \textsc{MedPriv}: Edge-side fusion defense} &
\multicolumn{2}{c}{\small (d) \textsc{CommPriv}: Edge-side fusion defense}
\end{tabular}
\vspace{-0.8em}
\caption{Privacy-utility trade-off curves: private evidence recall under Qwen2.5-72B/1.5B collaboration.}
\label{fig:qwen_privacy_utility_main}
\vspace{-0.5em}
\end{figure*}

\begin{figure*}[t]
\centering
\setlength{\tabcolsep}{0pt}
\begin{tabular}{@{}cccc@{}}
\includegraphics[width=0.249\textwidth]{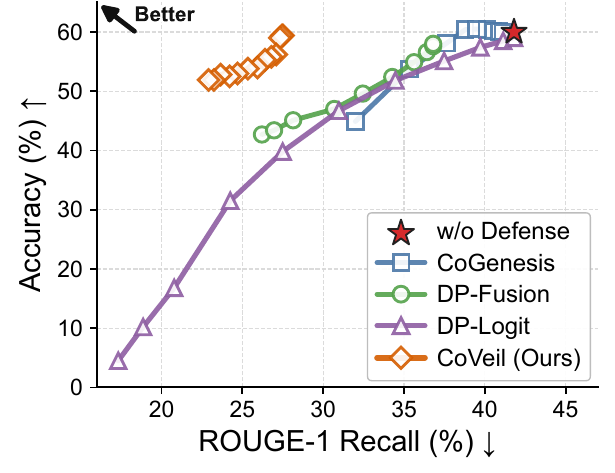} &
\includegraphics[width=0.249\textwidth]{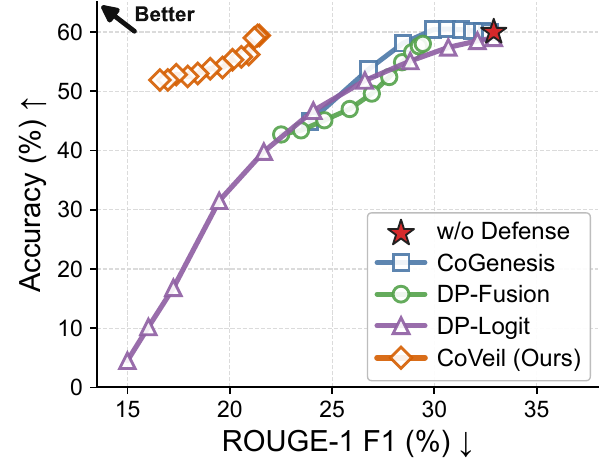} &
\includegraphics[width=0.249\textwidth]{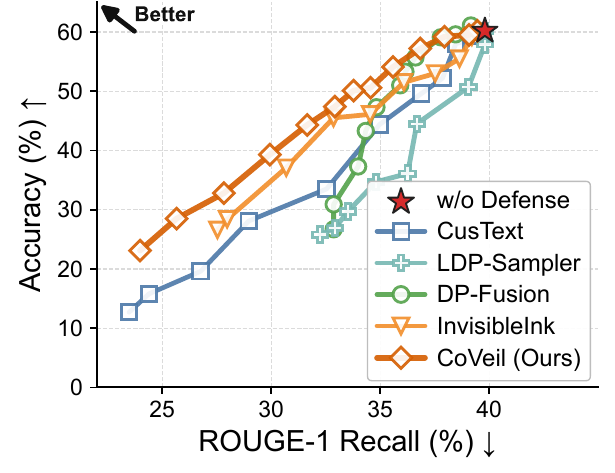} &
\includegraphics[width=0.249\textwidth]{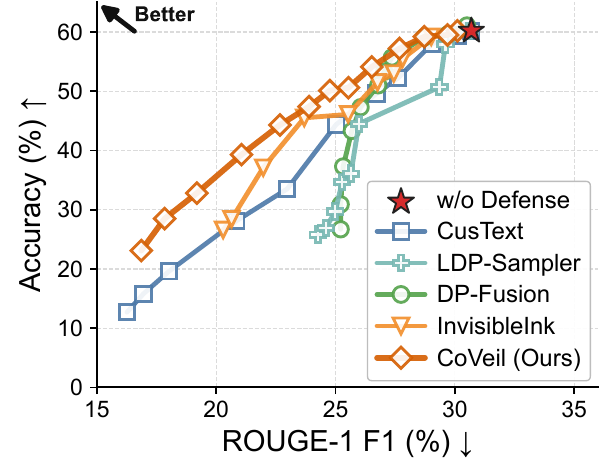} \\[-0.25em]
\multicolumn{2}{c}{\small (a) \textsc{MedPriv}: Cloud-side fusion defense} &
\multicolumn{2}{c}{\small (b) \textsc{MedPriv}: Edge-side fusion defense}
\end{tabular}
\vspace{-0.8em}
\caption{Privacy-utility trade-off curves: private context inversion under Qwen2.5-72B/1.5B collaboration.}
\label{fig:medqa_inversion_tradeoff}
\vspace{-0.5em}
\end{figure*}

\subsection{Edge-Side Fusion Defense}
In edge-side fusion, \textsc{CoVeil} selects a token from $\mathcal{C}_t=\operatorname{TopK}(p_{\mathrm{fuse}}^{(t)})$, rather than choosing an upload mask $m_t$ as in cloud-side fusion.
The selected token is appended to the shared autoregressive prefix at each decoding step.
For each candidate token $i\in\mathcal{C}_t$, the per-step objective is:
\begin{equation}
  \mathcal{J}_t(i)
  =
  -\mathcal{U}_t(i)
  +
  \lambda \cdot \mathcal{P}_t(i).
\end{equation}

\textbf{Utility function.}
Since edge-side fusion selects a token rather than an upload mask, its utility applies the probability-gap principle of \Cref{eq:cloud_side_utility} to each candidate token.
\textsc{CoVeil} scores each candidate by its fused-probability gap relative to the fused top token $y_t^\star$:
\begin{equation}
  \mathcal{U}_t(i)
  =
  p_{\mathrm{fuse}}^{(t)}(i)-p_{\mathrm{fuse}}^{(t)}(y_t^\star).
\end{equation}

\textbf{Privacy function.}
As in \Cref{eq:cloud_side_privacy}, edge-cloud discrepancy estimates private-context influence.
While cloud-side fusion penalizes uploaded deviations, edge-side fusion uses the signed difference for the selected token:
\begin{equation}
  \mathcal{P}_t(i)
  =
  p_S^{(t)}(i)-p_L^{(t)}(i).
\end{equation}
A positive value indicates stronger private-side support and thus higher leakage risk, whereas a negative value indicates stronger cloud-side support and lower leakage risk.

\textbf{Token selection rule.}
\textsc{CoVeil} selects the candidate that minimizes the per-step objective:
\begin{equation}
  i_t^\star
  =
  \arg\min_{i\in\mathcal{C}_t} \mathcal{J}_t(i).
\end{equation}

\section{Experiments}
\label{sec:experiments}

\subsection{Experimental Setup}

\noindent\textbf{Cloud-edge model setup.}
Experiments are conducted on two cloud-edge collaboration settings: Qwen2.5-72B~(LLM)/1.5B~(SLM)~\citep{qwen2024qwen2} and Llama-3.1-70B~(LLM)/Llama-3.2-3B~(SLM)~\citep{grattafiori2024llama}.
The corresponding tokenizer vocabulary sizes are 151,643 and 128,256, respectively.\footnote{Further implementation details are provided in \Cref{app:implementation_details,app:baseline_parameter_ranges}.}

\noindent\textbf{Datasets and evaluation metrics.}
We use \textsc{MedPriv} and \textsc{CommPriv} from \Cref{sec:signal_leakage}.
Privacy is evaluated by private evidence recall and private context inversion from \Cref{sec:coverage,sec:reconstruction}, and utility by answer accuracy.
Unless otherwise specified, the recall metrics in \Cref{sec:coverage} are reported at $K=100$, a compact head cutoff covering less than 0.1\% of either vocabulary.

\noindent\textbf{Comparison methods.}
Cloud-side fusion privacy defense baselines include CoGenesis~\citep{zhang2024cogenesis}, DP-Logit~\citep{ji2024less}, and DP-Fusion~\citep{thareja2026dp}.
Edge-side fusion privacy defense baselines include CusText~\citep{chen2023customized}, LDP-Sampler~\citep{ghoukasian2026locally}, DP-Fusion~\citep{thareja2026dp}, and InvisibleInk~\citep{vinod2026invisibleink}.
Several of these baselines provide method-specific formal differential privacy guarantees under their respective assumptions, whereas \textsc{CoVeil} provides empirical privacy improvements without a formal DP guarantee.
The trade-off weight $\lambda$ is not a formal privacy budget $\epsilon$; our comparisons are limited to empirical privacy-utility behavior under the common evaluation protocol.

\subsection{Main Results}

\paragraph{DecodeLeak Results.}
\Cref{tab:decodeleak_results} reports undefended leakage on \textsc{MedPriv}; the red stars in \Cref{fig:qwen_privacy_utility_main,fig:medqa_inversion_tradeoff} mark the corresponding no-defense points.
Both fusion modes expose private evidence through cloud-observed decoding signals.
Cloud-side fusion reaches 56.6\% Token-ER@100 and 41.1\% inversion ROUGE-1 Recall, while edge-side fusion reaches 43.0\% and 39.2\%, respectively.
These results show that undefended collaboration leaks private information at both the signal-exposure and context-inversion levels.

\paragraph{Cloud-Side Fusion Defense.}
As shown in \Cref{fig:qwen_privacy_utility_main}(a,b) and \Cref{fig:medqa_inversion_tradeoff}(a),\footnote{ROUGE1-ER@100, AUC@100, and LLaMA results are reported in \Cref{app:privacy_utility_tradeoffs,app:llama_results}.} \textsc{CoVeil} reduces Token-ER@100 by 75.7\% on \textsc{MedPriv} and 87.2\% on \textsc{CommPriv} with at most 0.2 percentage-point accuracy loss.
At matched accuracy (${\approx}59\%$), \textsc{CoVeil} achieves Token-ER@100 of 13.4\%, compared to 28.3\% for CoGenesis and 24.4\% for DP-Fusion.
\textsc{CoVeil} outperforms these baselines by evaluating the utility-privacy trade-off at each token position, whereas fixed upload constraints in CoGenesis and DP-Fusion leave leaky positions exposed and uniform perturbation in DP-Logit reduces leakage only with sharp accuracy drops.

\paragraph{Edge-Side Fusion Defense.}
\Cref{fig:qwen_privacy_utility_main}(c,d) reports private-evidence recall, and \Cref{fig:medqa_inversion_tradeoff}(b) reports private context inversion under edge-side fusion.
Edge-side fusion exposes stronger privacy leakage because each synchronized token is appended to the cloud-observed prefix and depends on the private-conditioned edge distribution.
\textsc{CoVeil} reduces Token-ER@100 from 43.0\% to 40.4\% and Span-ER@100 from 28.3\% to 26.9\% on \textsc{MedPriv}, and from 36.6\% to 35.5\% and 27.5\% to 27.0\% on \textsc{CommPriv}.
At matched accuracy, \textsc{CoVeil} achieves lower Token-ER@100 than InvisibleInk, DP-Fusion, and CusText; at matched leakage, it achieves 7.2\% higher accuracy than InvisibleInk.
This advantage comes from utility-aware token selection: \textsc{CoVeil} avoids appending tokens strongly supported by the private-side model unless they provide sufficient utility, whereas token-side baselines rely on preset sampling or perturbation rules.

\begin{table}[t]
\centering
\small
\setlength{\tabcolsep}{2.2pt}
\resizebox{\columnwidth}{!}{%
\begin{tabular}{@{}lc@{\hspace{4pt}}cccc@{\hspace{4pt}}cc@{}}
\toprule
\multirow{2}{*}{\textbf{Fusion}} &
\multirow{2}{*}{\textbf{Acc}} &
\multicolumn{4}{c}{\textbf{Private Evidence Recall} (@100)} &
\multicolumn{2}{c}{\textbf{Private Context Inversion}} \\
\cmidrule(lr){3-6}\cmidrule(lr){7-8}
& & \textbf{Token-ER} & \textbf{ROUGE1-ER} & \textbf{Span-ER} & \textbf{AUC}
  & \textbf{ROUGE-1 Recall} & \textbf{ROUGE-1 F1} \\
\midrule
Cloud-side & 59.3 & 56.6 & 48.3 & 28.3 & 46.7 & 41.1 & 32.7 \\
Edge-side  & 59.3 & 43.0 & 34.6 & 28.3 & 35.0 & 39.2 & 30.3 \\
\bottomrule
\end{tabular}
}
\vspace{-0.5em}
\caption{\textsc{DecodeLeak} results on \textsc{MedPriv} under undefended Qwen2.5-72B/1.5B collaboration.}
\vspace{-1.0em}
\label{tab:decodeleak_results}
\end{table}

\subsection{Sensitivity Analysis}
\label{sec:step_level_mechanism}
We analyze per-step decisions of \textsc{CoVeil}: cloud-side fusion chooses which SLM token probabilities to upload, while edge-side fusion chooses which token to append to the shared autoregressive prefix.
We sweep $\lambda$ to connect these decisions with the observed privacy-utility trends.

\begin{figure}[t]
\centering
\includegraphics[width=0.85\columnwidth]{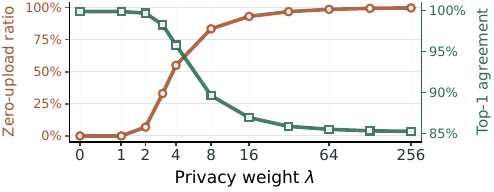}\\[-0.0em]

\small (a) Cloud-side fusion: adaptive upload

\includegraphics[width=0.85\columnwidth]{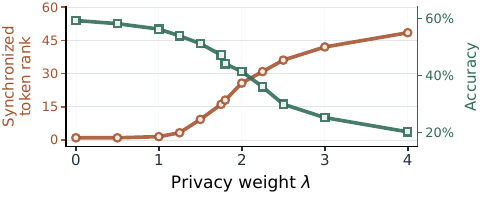}\\[-0.0em]
\small (b) Edge-side fusion: token selection
\vspace{-0.5em}
\caption{Effect of the privacy weight $\lambda$ on per-step \textsc{CoVeil} decisions on \textsc{MedPriv}.}
\vspace{-1.0em}
\label{fig:step_level_mechanism}
\end{figure}

\begin{figure}[t]
\centering
\includegraphics[width=0.9\columnwidth]{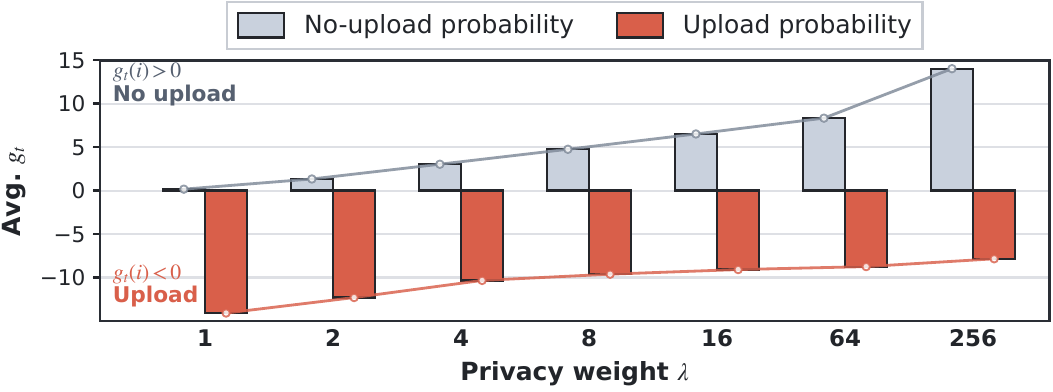}
\vspace{-0.5em}
\caption{Cloud-side upload decisions on \textsc{MedPriv}: candidate SLM token probabilities are uploaded when $g_t(i)<0$ and suppressed when $g_t(i)\ge0$.}
\vspace{-0.5em}
\label{fig:lm_upload_gradient_trigger}
\end{figure}

\begin{figure}[t]
\centering
\includegraphics[width=0.9\columnwidth]{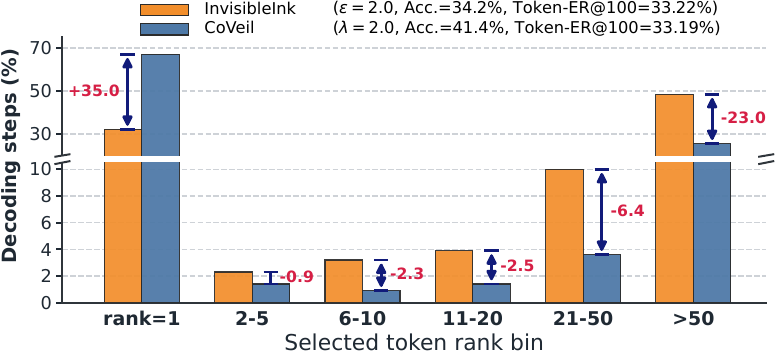}
\vspace{-0.5em}
\caption{Edge-side selected-token rank distribution at matched Token-ER@100 on \textsc{MedPriv}; ranks are computed under $p_{\mathrm{fuse}}^{(t)}$.}
\label{fig:sm_rank_bin_invisibleink}
\vspace{-1.0em}
\end{figure}

\paragraph{Effect of $\lambda$ in cloud-side defense.}
\Cref{fig:step_level_mechanism}(a) analyzes how the privacy weight $\lambda$ changes cloud-side upload behavior.
The left y-axis reports the zero-upload ratio, the fraction of decoding steps where \textsc{CoVeil} uploads no SLM token probability; the right y-axis reports top-1 agreement with the original full-fusion choice.
As $\lambda$ increases, the zero-upload ratio rises because the objective assigns more weight to privacy cost.
At $\lambda=8$, 83.3\% of steps upload no SLM probabilities, while top-1 agreement remains 89.6\%.
This shows that many SLM token probabilities can be suppressed without changing the full-fusion top choice in most steps, reducing privacy exposure.
\Cref{fig:lm_upload_gradient_trigger} further illustrates the per-token upload rule: candidates with $g_t(i)<0$ reduce the objective and are uploaded, while candidates with $g_t(i)\ge 0$ are not.

\paragraph{Effect of $\lambda$ in edge-side defense.}
\Cref{fig:step_level_mechanism}(b) analyzes how privacy weight $\lambda$ changes edge-side token selection.
The left y-axis reports selected-token rank under $p_{\mathrm{fuse}}^{(t)}$, and the right y-axis reports accuracy.
Edge-side decisions are more constrained because each step must append one token to shared autoregressive prefix.
For $\lambda \in [0,1]$, the average rank increases only from 1.0 to 1.5 while accuracy decreases from 59.3\% to 56.3\%; at $\lambda=4.0$, the rank reaches 48.4 and accuracy drops to 20.2\%.
At matched leakage (\Cref{fig:sm_rank_bin_invisibleink}; Token-ER@100: 33.19\% vs.\ 33.22\%), \textsc{CoVeil} retains 67.1\% of steps at rank~1 versus 32.1\% in InvisibleInk, and reduces selections below rank 50 from 48.5\% to 25.5\%.
This shows that \textsc{CoVeil} preserves the fused top token more often at matched leakage, consistent with its utility-aware token selection rule.

\section{Related Work}

Large-small language model collaboration improves generation quality and inference efficiency~\citep{li2023contrastive,leviathan2023fast,liu2024proxy}, typically with both models sharing the same input context.
Cloud-edge collaborative decoding instead keeps private context on the device and exposes only cloud-observed decoding signals, including uploaded SLM probabilities or synchronized tokens~\citep{zhang2024cogenesis,tian2026floe,she2025token}.
These systems mainly target utility, cost, or latency, leaving privacy leakage from cloud-observed decoding signals underexplored.

Privacy attacks show that outputs and next-token distributions can reveal hidden inputs through language-model inversion~\citep{morris2024language,skapars2025gpt}, prompt extraction~\citep{qu2025prompt,luo2025prompt}, or probability analysis~\citep{finlaysonlogits,flemings2025estimating}.
Typical inversion attacks require supervised training on public data~\citep{nazir2025better,zhang2024extracting} or white-box gradient access~\citep{skapars2025gpt}.
However, these studies do not analyze cloud-edge collaborative decoding, where the cloud observes per-step signals.
\textsc{DecodeLeak} instead measures private evidence recall and performs private context inversion from the signal exposure sequence.

Existing defenses address exposed signals through two routes.
For uploaded distributions, CoGenesis sparsifies SLM probabilities~\citep{zhang2024cogenesis}, DP-Logit adds noise~\citep{ji2024less}, and DP-Fusion calibrates the private-conditioned distribution toward a public baseline~\citep{thareja2026dp}.
For synchronized tokens, CusText sanitizes tokens~\citep{chen2023customized}, LDP-Sampler samples from a public-anchor distribution~\citep{ghoukasian2026locally}, DP-Fusion samples from its calibrated distribution~\citep{thareja2026dp}, and InvisibleInk clips the SLM--LLM residual before private sampling~\citep{vinod2026invisibleink}.
In contrast, \textsc{CoVeil} makes per-step, fusion-mode-specific decisions by optimizing the utility-privacy trade-off.

\section{Conclusion}

We analyzed privacy leakage in cloud-edge collaborative decoding, where private context remains on the edge but the cloud observes per-step decoding signals.
To support this analysis, we introduced \textsc{DecodeLeak} with two QA benchmarks, showing that uploaded SLM probabilities and synchronized tokens can expose private evidence and enable private context inversion.
We further proposed \textsc{CoVeil}, which makes per-step upload and synchronization decisions under a utility-privacy objective.
Experiments show that \textsc{CoVeil} improves the privacy-utility trade-off in both fusion modes, highlighting that cloud-observed decoding signals should be treated as privacy-sensitive interfaces.

\section*{Limitations}

Our experiments consider cloud-edge model pairs from the same model family with a shared tokenizer vocabulary.
This setting is aligned with probability-level fusion, where next-token probabilities are compared or combined over a common token space.
Extending the framework to heterogeneous tokenizers would require explicit tokenizer or probability alignment before fusion, which we leave to future work.
For private context inversion, we adopt a prompt-based attacker that provides the cloud LLM with the public query and accumulated signal exposure sets.
We do not systematically optimize inversion prompts or train a separate inversion model.
This choice keeps the evaluation simple and reproducible, but stronger inversion strategies may affect the measured reconstruction quality.

\section*{Acknowledgments}

This work was supported by the Presidential Young Scholar Scheme (Project No. P0056638), RIFL (Project No. 4-CG00), and RIAIoT (Project No. P0059914) at The Hong Kong Polytechnic University, as well as by the Artificial Intelligence Research Centre at the PolyU--Daya Bay Technology and Innovation Research Institute.

\bibliography{reference}

\clearpage

\appendix

\section{Appendix}
\setcounter{table}{0}
\renewcommand{\thetable}{A\arabic{table}}

\subsection{Implementation Details}
\label{app:implementation_details}

\subsubsection{Base Model Setup}
\label{app:base_model_setup}

We evaluate two cloud and edge model pairs.
For the Qwen family, we use Qwen2.5-1.5B as the edge model and Qwen2.5-72B as the cloud model.
For the LLaMA family, we use Llama-3.2-3B as the edge model and Llama-3.1-70B-Instruct as the cloud model.
In each pair, the smaller model runs on the edge and conditions on both the public query $x_{\mathrm{pub}}$ and the private context $x_{\mathrm{priv}}$.
The larger model runs on the cloud and conditions only on $x_{\mathrm{pub}}$ and the shared generated prefix.
Within each model family, the edge and cloud models use the same tokenizer vocabulary.
The Qwen pair uses a vocabulary of 151,643 tokens, and the LLaMA pair uses a vocabulary of 128,256 tokens.
This shared vocabulary allows their next-token probability distributions to be compared and fused over the same token space.
We therefore do not use tokenizer remapping or cross-vocabulary probability alignment in our experiments.

\subsubsection{Collaborative Decoding Setup}
\label{app:collaborative_decoding_setup}

We use the same collaborative decoding configuration across all model families, datasets, fusion modes, and defense settings.
After parameter analysis, we fix the fusion weight to $\alpha=0.3$.
At each decoding step, the fused next-token distribution is computed by weighted interpolation of the cloud and edge next-token distributions, following \Cref{sec:setting}.
All experiments use greedy decoding with temperature 0 to ensure deterministic outputs and reproducibility.
We set \texttt{max\_new\_tokens} to 256 for all runs.
This limit covers the answer and rationale format, and generation stops earlier when an end-of-sequence token is produced.
For each example, all methods receive the same public query $x_{\mathrm{pub}}$, answer options $\mathcal{A}$, and output instruction.
Both fusion modes maintain the same shared autoregressive prefix $y_{<t}$.
In cloud-side fusion, the cloud fuses $p_L^{(t)}$ and $p_S^{(t)}$ and appends $y_t^{\mathrm{fuse}}$ to $y_{<t}$.
In edge-side fusion, the edge selects $y_t^{\mathrm{fuse}}$ from the fused distribution, sends it to the cloud, and both sides append it to $y_{<t}$.

\subsubsection{\textsc{CoVeil} Parameter Settings}
\label{app:coveil_parameter_settings}

The main tunable parameter in \textsc{CoVeil} is the privacy weight $\lambda$ in the per-step objective.
For cloud-side fusion, the main experiments use
\[
\lambda \in \{0, 0.5, 1, 1.5, 2, 2.5, 3, 4, 6, 8, 12, 16\}.
\]
For edge-side fusion, where utility drops more sharply as $\lambda$ increases, the main experiments use
\[
\lambda \in \{0, 0.5, 1, 1.25, 1.5, 1.75, 1.8, 2, 2.25, 2.5, 3, 4\},
\]
for both \textsc{MedPriv} and \textsc{CommPriv}.
Larger $\lambda$ values in the mechanism analysis are used only to study behavior beyond the main operating range.

\subsubsection{Defense Baseline Adaptations and Parameter Ranges}
\label{app:baseline_parameter_ranges}
\label{app:cloud_side_fusion_defense_baselines}
\label{app:edge_side_fusion_defense_baselines}

We group comparison methods by the cloud-observed signal they protect.
Cloud-side baselines act on the uploaded SLM next-token probabilities, while edge-side baselines act on the token appended to the shared prefix.
Except for the swept privacy-control parameter, we follow the original method settings whenever applicable.

\paragraph{Cloud-side baselines.}
\textbf{CoGenesis} uploads only the top-$B$ SLM probabilities at each step~\citep{zhang2024cogenesis}.
\textbf{DP-Logit} adds Gaussian noise to the uploaded SLM signal before fusion~\citep{ji2024less}.
\textbf{DP-Fusion} calibrates the private-conditioned SLM distribution toward a public baseline before upload~\citep{thareja2026dp}.

\paragraph{Edge-side baselines.}
\textbf{CusText} sanitizes the token selected from the private-conditioned decoding distribution before prefix update~\citep{chen2023customized}.
\textbf{LDP-Sampler} uses $p_S^{(t)}$ as the private distribution and $p_L^{(t)}$ as the public anchor~\citep{ghoukasian2026locally}.
\textbf{DP-Fusion} samples from the public-baseline-calibrated private distribution~\citep{thareja2026dp}.
\textbf{InvisibleInk} clips the SLM--LLM logit residual before sampling one token for prefix update~\citep{vinod2026invisibleink}.

\begin{center}
\centering
\scriptsize
\setlength{\tabcolsep}{2pt}
\renewcommand{\arraystretch}{0.98}
\resizebox{\columnwidth}{!}{%
\begin{tabular}{@{}ll@{}}
\toprule
\textbf{Method} & \textbf{Swept parameter} \\
\midrule
\multicolumn{2}{@{}l}{\textit{Cloud-side fusion}} \\
CoGenesis & $B \in \{1, 2, 4, 8, 10, 20, 30, 50, 100\}$ \\
DP-Logit & $\epsilon \in \{0.01, 0.05, 0.1, 0.5, 1, 2, 4, 8, 16, 32, 64\}$ \\
DP-Fusion & $\epsilon \in \{0, 0.1, 0.5, 1, 2, 4, 8, 16, 32, 64\}$ \\
\midrule
\multicolumn{2}{@{}l}{\textit{Edge-side fusion}} \\
CusText & $\epsilon \in \{1, 2, 4, 6, 8, 10, 12, 14, 16, 24, 32\}$ \\
LDP-Sampler & $\epsilon \in \{0.1, 0.5, 1, 2, 4, 8, 32, 64\}$ \\
DP-Fusion & $\epsilon \in \{0, 0.1, 0.5, 1, 2, 3, 4, 8, 16, 32, 64\}$ \\
InvisibleInk & $\epsilon \in \{1, 2, 4, 8, 16, 32, 64, 128\}$ \\
\bottomrule
\end{tabular}
}
\vspace{0.25em}

\parbox{0.98\columnwidth}{\centering
\textbf{Swept parameters for comparison methods.} \textsc{CoVeil} uses the $\lambda$ grids in \Cref{app:coveil_parameter_settings}.}
\end{center}

\begin{figure*}[t]
\centering
\setlength{\tabcolsep}{0pt}
\begin{tabular}{@{}cccc@{}}
\includegraphics[width=0.249\textwidth]{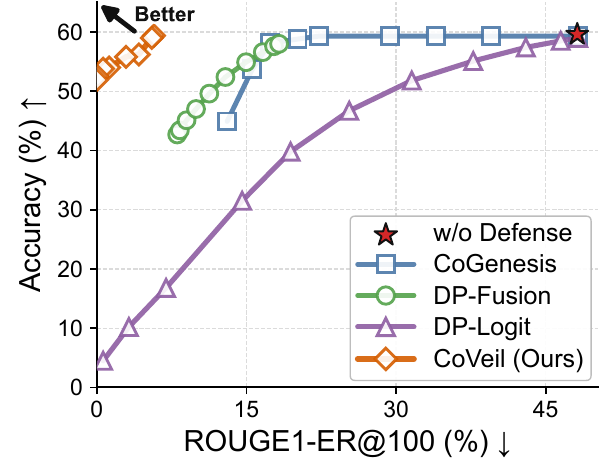} &
\includegraphics[width=0.249\textwidth]{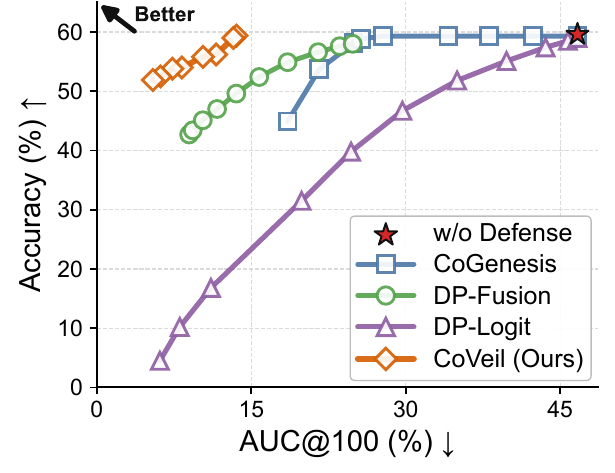} &
\includegraphics[width=0.249\textwidth]{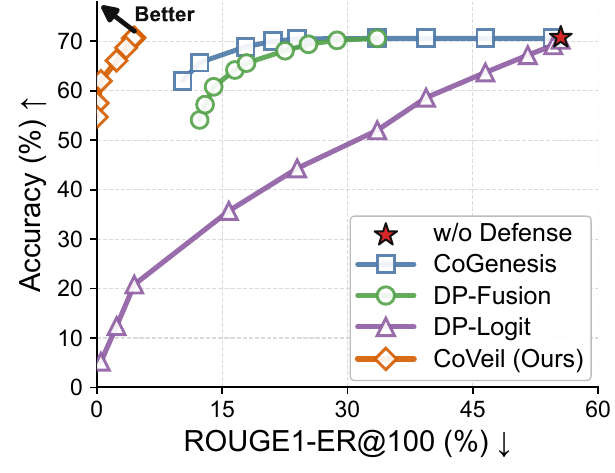} &
\includegraphics[width=0.249\textwidth]{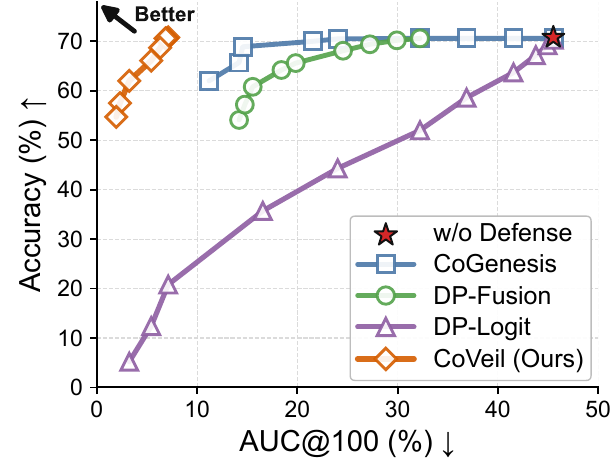} \\[-0.35em]
\multicolumn{2}{c}{\small (a) \textsc{MedPriv}: Cloud-side fusion defense} &
\multicolumn{2}{c}{\small (b) \textsc{CommPriv}: Cloud-side fusion defense} \\[-0.2em]
\includegraphics[width=0.249\textwidth]{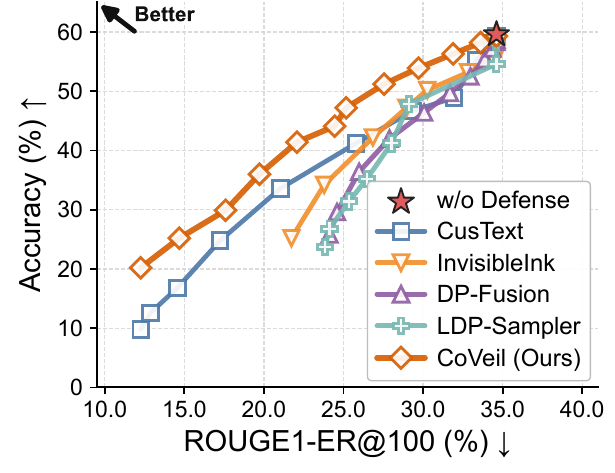} &
\includegraphics[width=0.249\textwidth]{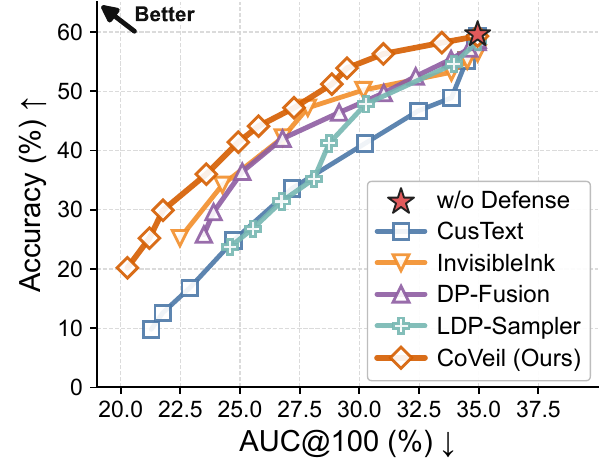} &
\includegraphics[width=0.249\textwidth]{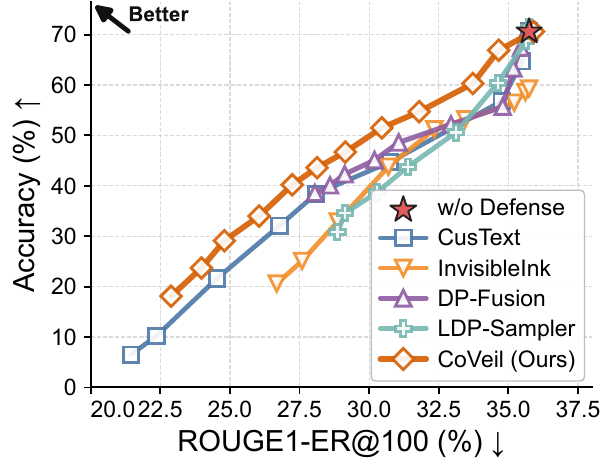} &
\includegraphics[width=0.249\textwidth]{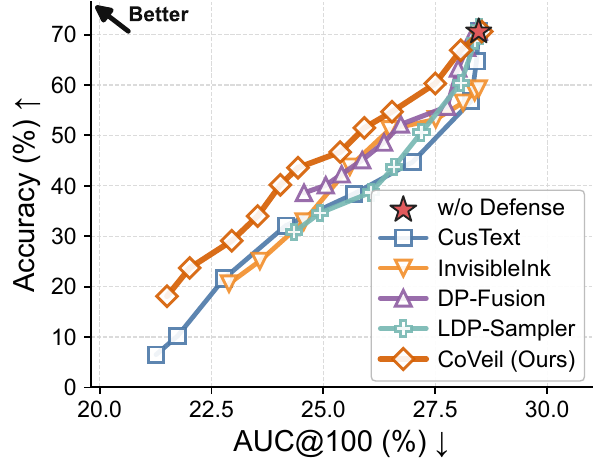} \\[-0.35em]
\multicolumn{2}{c}{\small (c) \textsc{MedPriv}: Edge-side fusion defense} &
\multicolumn{2}{c}{\small (d) \textsc{CommPriv}: Edge-side fusion defense}
\end{tabular}
\vspace{-0.5em}
\caption{Additional privacy-utility trade-off curves for Qwen2.5-72B/1.5B cloud-edge collaboration on ROUGE1-ER@100 and AUC@100.}
\label{fig:qwen_privacy_utility_appendix}
\end{figure*}

\begin{figure*}[t]
\centering
\setlength{\tabcolsep}{0pt}
\begin{tabular}{@{}cccc@{}}
\includegraphics[width=0.249\textwidth]{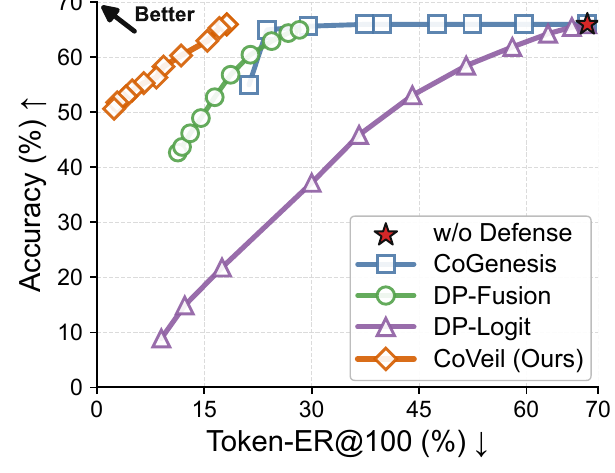} &
\includegraphics[width=0.249\textwidth]{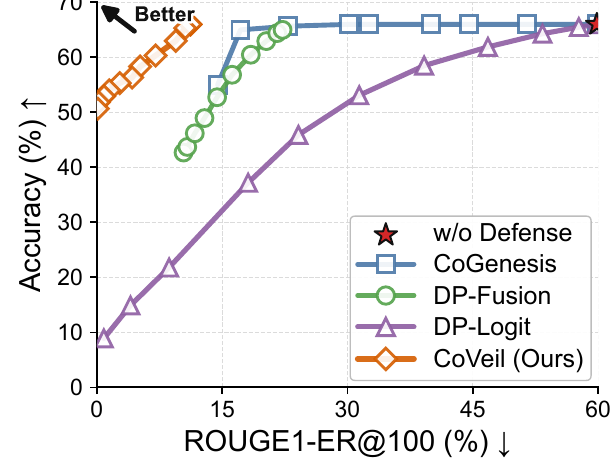} &
\includegraphics[width=0.249\textwidth]{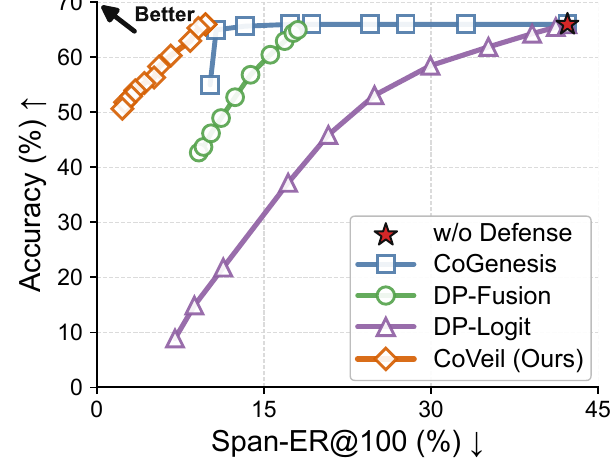} &
\includegraphics[width=0.249\textwidth]{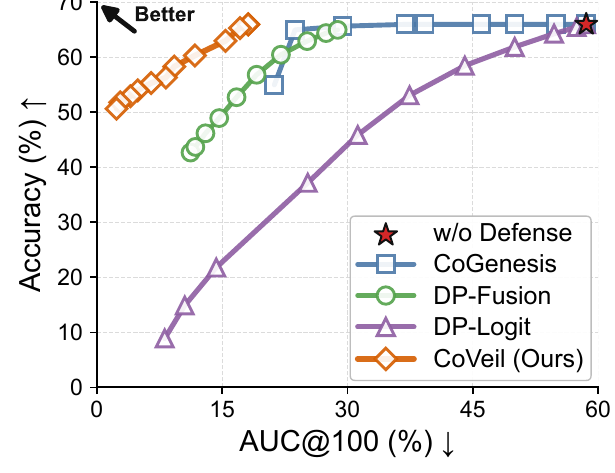} \\[-0.35em]
\multicolumn{4}{c}{\small (a) Cloud-side fusion defense} \\[-0.2em]
\includegraphics[width=0.249\textwidth]{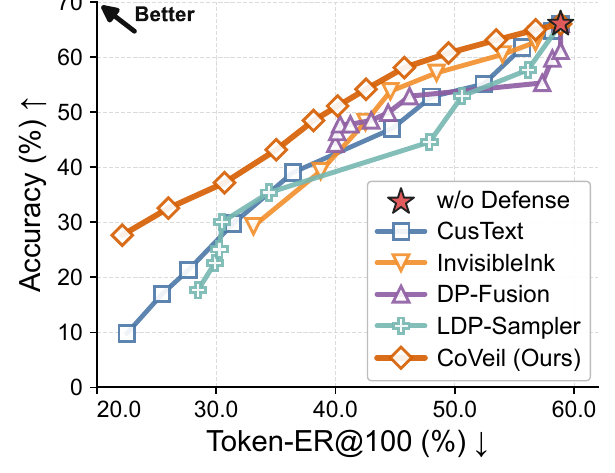} &
\includegraphics[width=0.249\textwidth]{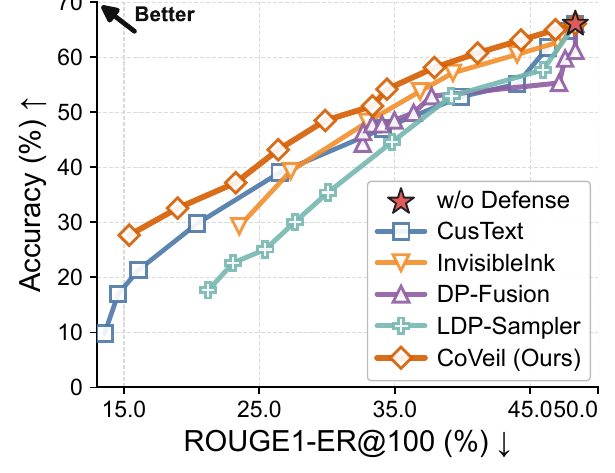} &
\includegraphics[width=0.249\textwidth]{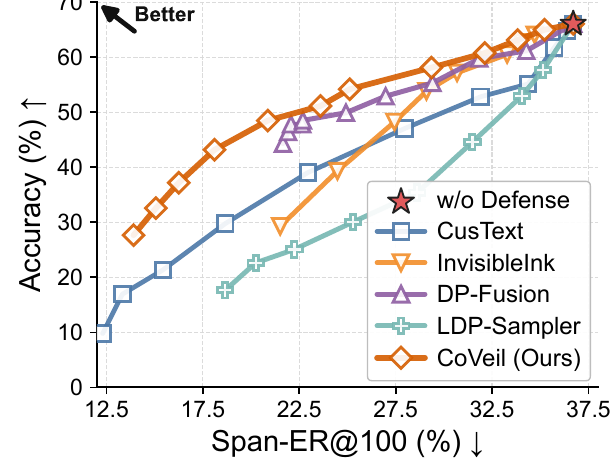} &
\includegraphics[width=0.249\textwidth]{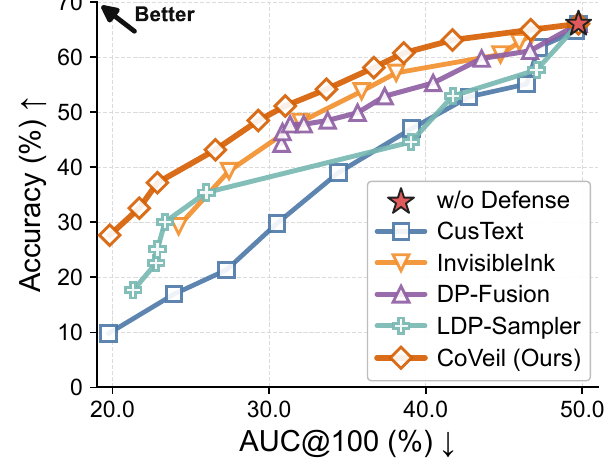} \\[-0.35em]
\multicolumn{4}{c}{\small (b) Edge-side fusion defense}
\end{tabular}
\vspace{-0.5em}
\caption{Cloud-side and edge-side privacy-utility trade-off curves for LLaMA-3.2-3B/LLaMA-3.1-70B cloud-edge collaboration on \textsc{MedPriv}. These appendix results correspond to the main-text privacy-utility evaluation and report private evidence recall with Token-ER@100, ROUGE1-ER@100, Span-ER@100, and AUC@100.}
\label{fig:llama_medpriv_privacy_utility_appendix}
\end{figure*}

\subsection{Additional Experimental Results}
\label{app:additional_experimental_results}
\label{app:privacy_utility_tradeoffs}

\subsubsection{Qwen Results}
\label{app:qwen_results}

\Cref{tab:qwen_undefended_baselines} isolates the utility gain from undefended cloud-edge collaboration.
On \textsc{MedPriv}, collaboration improves accuracy from the stronger single-model setting of 44.5 to 59.6, a gain of 15.1 points.
On \textsc{CommPriv}, the gain is similar: accuracy increases from 55.9 to 70.8.
These gains confirm that the private context contains answer-critical information that is not recoverable from the public query alone, while cloud-edge collaborative decoding can use the edge SLM next-token distribution $p_S^{(t)}$, which conditions on $x_{\mathrm{pub}}$ and $x_{\mathrm{priv}}$, to improve answer selection.

\Cref{fig:qwen_privacy_utility_appendix} reports the ROUGE1-ER@100 and AUC@100 counterparts of \Cref{fig:qwen_privacy_utility_main}.
The trends are consistent with the main results: \textsc{CoVeil} achieves lower private evidence recall at comparable accuracy under cloud-side fusion, while edge-side fusion shows a tighter trade-off because the synchronized token is appended to the shared autoregressive prefix at each decoding step.
The agreement across Token-ER, Span-ER, ROUGE1-ER, and AUC indicates that the observed privacy reduction is not an artifact of a single private evidence recall metric.
Since the cloud-side and edge-side variants have the same undefended task accuracy, we collapse them into one collaboration row and leave fusion-mode-specific private evidence recall scores to the \textsc{CoVeil} sweeps in \Cref{tab:qwen_cloudside_coveil_sweep,tab:qwen_edgeside_coveil_sweep}.

\Cref{tab:qwen_cloudside_coveil_sweep,tab:qwen_edgeside_coveil_sweep} make the role of the privacy weight explicit.
For cloud-side fusion, increasing $\lambda$ steadily suppresses recall with a gradual utility cost: on \textsc{MedPriv}, Token-ER@100 decreases from 0.1374 at $\lambda=0$ to 0.0582 at $\lambda=16$, while accuracy decreases from 59.4 to 51.9.
For edge-side fusion, the trade-off is steeper because the system must still append one synchronized token to the shared autoregressive prefix at each decoding step; on \textsc{MedPriv}, Token-ER@100 decreases from 0.4301 to 0.2560 as $\lambda$ increases from 0 to 4, but accuracy drops from 59.3 to 20.2.
This difference explains why cloud-side adaptive upload admits stronger leakage reduction at near-constant utility, whereas edge-side token selection yields a tighter utility-privacy trade-off.

\paragraph{Top-$K$ candidate sensitivity.}
\label{app:topk_candidate_sensitivity}
\Cref{tab:topk_candidate_sensitivity} evaluates the candidate size used by the cloud-side upload mask on \textsc{MedPriv} with $\lambda=1.0$.
Here, $K$ denotes the candidate set size for upload optimization, while @100 denotes the exposure cutoff used by the leakage metrics.
The results are stable once $K\ge 50$: increasing the candidate set does not change accuracy or leakage metrics, suggesting that larger candidate sets mostly add low-probability positions without materially affecting the adaptive upload decisions.

\begin{table}[t]
\centering
\scriptsize
\setlength{\tabcolsep}{4pt}
\renewcommand{\arraystretch}{0.95}
\resizebox{\columnwidth}{!}{%
\begin{tabular}{@{}lccccc@{}}
\toprule
\textbf{Setting} & \textbf{Acc.} & \textbf{Token-ER@100} & \textbf{ROUGE1-ER@100} & \textbf{Span-ER@100} & \textbf{AUC@100} \\
\midrule
w/o \textsc{CoVeil} & 59.60 & 0.5655 & 0.4817 & 0.2823 & 0.4668 \\
\textsc{CoVeil}, $K=10$ & 58.80 & 0.1320 & 0.0542 & 0.0802 & 0.1299 \\
\textsc{CoVeil}, $K=50$ & 59.00 & 0.1344 & 0.0551 & 0.0822 & 0.1325 \\
\textsc{CoVeil}, $K=100$ & 59.00 & 0.1344 & 0.0551 & 0.0822 & 0.1325 \\
\textsc{CoVeil}, $K=500$ & 59.00 & 0.1344 & 0.0551 & 0.0822 & 0.1325 \\
\textsc{CoVeil}, $K=1000$ & 59.00 & 0.1344 & 0.0551 & 0.0822 & 0.1325 \\
\bottomrule
\end{tabular}
}
\caption{Sensitivity to the upload candidate size $K$ under cloud-side fusion on \textsc{MedPriv}. \textsc{CoVeil} uses $\lambda=1.0$. Privacy metrics are raw scores.}
\label{tab:topk_candidate_sensitivity}
\end{table}

\begin{table}[t]
\centering
\small
\setlength{\tabcolsep}{5pt}
\renewcommand{\arraystretch}{0.96}
\resizebox{\columnwidth}{!}{%
\begin{tabular}{@{}llc@{}}
\toprule
\textbf{Data} & \textbf{Setting} & \textbf{Acc.} \\
\midrule
\multirow{4}{*}{\textsc{MedPriv}}
& Cloud LLM (w/o private context) & 33.9 \\
& Edge SLM (w/ private context) & 44.5 \\
& \textbf{Cloud-edge collaboration (w/ private context)} & \textbf{59.6} \\
& Gain over best single model & +15.1 \\
\midrule
\multirow{4}{*}{\textsc{CommPriv}}
& Cloud LLM (w/o private context) & 35.6 \\
& Edge SLM (w/ private context) & 55.9 \\
& \textbf{Cloud-edge collaboration (w/ private context)} & \textbf{70.8} \\
& Gain over best single model & +14.9 \\
\bottomrule
\end{tabular}
}
\caption{Undefended Qwen2.5-72B/1.5B collaboration accuracy. Here, w/o private context denotes the cloud LLM using only the public query, while w/ private context denotes settings where the edge SLM conditions on the private context. The gain is measured against the stronger of the two non-collaborative settings for each benchmark.}
\label{tab:qwen_undefended_baselines}
\end{table}

\begin{table}[t]
\centering
\scriptsize
\setlength{\tabcolsep}{4pt}
\renewcommand{\arraystretch}{0.95}
\resizebox{\columnwidth}{!}{%
\begin{tabular}{@{}lcccccc@{}}
\toprule
\textbf{Data} & \textbf{$\lambda$} & \textbf{Acc.} & \textbf{Token-ER@100} & \textbf{ROUGE1-ER@100} & \textbf{Span-ER@100} & \textbf{AUC@100} \\
\midrule
\multirow{12}{*}{\textsc{MedPriv}}
& 0.0 & 59.4 & 0.1374 & 0.0573 & 0.0839 & 0.1356 \\
& 0.5 & 59.4 & 0.1374 & 0.0573 & 0.0839 & 0.1356 \\
& 1.0 & 59.0 & 0.1345 & 0.0551 & 0.0823 & 0.1326 \\
& 1.5 & 56.2 & 0.1178 & 0.0420 & 0.0745 & 0.1159 \\
& 2.0 & 55.8 & 0.1049 & 0.0293 & 0.0685 & 0.1030 \\
& 2.5 & 55.3 & 0.0922 & 0.0183 & 0.0625 & 0.0904 \\
& 3.0 & 54.0 & 0.0845 & 0.0124 & 0.0596 & 0.0826 \\
& 4.0 & 53.8 & 0.0753 & 0.0062 & 0.0553 & 0.0734 \\
& 6.0 & 53.0 & 0.0676 & 0.0023 & 0.0512 & 0.0656 \\
& 8.0 & 52.5 & 0.0637 & 0.0008 & 0.0494 & 0.0617 \\
& 12.0 & 52.8 & 0.0600 & 0.0003 & 0.0480 & 0.0581 \\
& 16.0 & 51.9 & 0.0582 & 0.0002 & 0.0474 & 0.0562 \\
\midrule
\multirow{12}{*}{\textsc{CommPriv}}
& 0.0 & 70.8 & 0.0717 & 0.0448 & 0.0570 & 0.0712 \\
& 0.5 & 70.6 & 0.0715 & 0.0445 & 0.0554 & 0.0701 \\
& 1.0 & 70.6 & 0.0690 & 0.0428 & 0.0546 & 0.0686 \\
& 1.5 & 68.7 & 0.0637 & 0.0350 & 0.0499 & 0.0632 \\
& 2.0 & 66.1 & 0.0547 & 0.0236 & 0.0447 & 0.0542 \\
& 2.5 & 65.1 & 0.0479 & 0.0166 & 0.0399 & 0.0474 \\
& 3.0 & 64.4 & 0.0412 & 0.0103 & 0.0346 & 0.0407 \\
& 4.0 & 62.0 & 0.0328 & 0.0048 & 0.0280 & 0.0323 \\
& 6.0 & 59.7 & 0.0259 & 0.0013 & 0.0227 & 0.0254 \\
& 8.0 & 57.5 & 0.0235 & 0.0010 & 0.0211 & 0.0230 \\
& 12.0 & 56.5 & 0.0212 & 0.0002 & 0.0191 & 0.0207 \\
& 16.0 & 54.7 & 0.0198 & 0.0001 & 0.0184 & 0.0193 \\
\bottomrule
\end{tabular}
}
\caption{Cloud-side privacy-weight sweep for \textsc{CoVeil} on Qwen2.5-72B/1.5B collaboration. Privacy metrics are raw scores.}
\label{tab:qwen_cloudside_coveil_sweep}
\end{table}

\begin{table}[t]
\centering
\scriptsize
\setlength{\tabcolsep}{4pt}
\renewcommand{\arraystretch}{0.95}
\resizebox{\columnwidth}{!}{%
\begin{tabular}{@{}lcccccc@{}}
\toprule
\textbf{Data} & \textbf{$\lambda$} & \textbf{Acc.} & \textbf{Token-ER@100} & \textbf{ROUGE1-ER@100} & \textbf{Span-ER@100} & \textbf{AUC@100} \\
\midrule
\multirow{12}{*}{\textsc{MedPriv}}
& 0.0 & 59.3 & 0.4301 & 0.3461 & 0.2829 & 0.3495 \\
& 0.5 & 58.2 & 0.4201 & 0.3362 & 0.2756 & 0.3345 \\
& 1.0 & 56.3 & 0.4045 & 0.3189 & 0.2688 & 0.3100 \\
& 1.25 & 53.9 & 0.3856 & 0.2971 & 0.2605 & 0.2948 \\
& 1.5 & 51.2 & 0.3680 & 0.2755 & 0.2469 & 0.2854 \\
& 1.75 & 47.2 & 0.3528 & 0.2517 & 0.2262 & 0.2706 \\
& 1.8 & 44.1 & 0.3415 & 0.2445 & 0.2187 & 0.2577 \\
& 2.0 & 41.4 & 0.3319 & 0.2207 & 0.2054 & 0.2493 \\
& 2.25 & 36.0 & 0.3171 & 0.1972 & 0.1918 & 0.2359 \\
& 2.5 & 29.9 & 0.2966 & 0.1759 & 0.1828 & 0.2177 \\
& 3.0 & 25.2 & 0.2743 & 0.1468 & 0.1769 & 0.2120 \\
& 4.0 & 20.2 & 0.2560 & 0.1225 & 0.1713 & 0.2028 \\
\midrule
\multirow{12}{*}{\textsc{CommPriv}}
& 0.0 & 70.6 & 0.3666 & 0.3590 & 0.2753 & 0.2854 \\
& 0.5 & 66.9 & 0.3550 & 0.3465 & 0.2703 & 0.2807 \\
& 1.0 & 60.3 & 0.3441 & 0.3372 & 0.2643 & 0.2750 \\
& 1.25 & 54.7 & 0.3376 & 0.3180 & 0.2579 & 0.2653 \\
& 1.5 & 51.5 & 0.3301 & 0.3045 & 0.2516 & 0.2591 \\
& 1.75 & 46.7 & 0.3207 & 0.2914 & 0.2456 & 0.2538 \\
& 1.8 & 43.6 & 0.3154 & 0.2813 & 0.2400 & 0.2443 \\
& 2.0 & 40.2 & 0.3102 & 0.2724 & 0.2351 & 0.2404 \\
& 2.25 & 34.0 & 0.2920 & 0.2605 & 0.2314 & 0.2353 \\
& 2.5 & 29.1 & 0.2855 & 0.2480 & 0.2237 & 0.2295 \\
& 3.0 & 23.7 & 0.2747 & 0.2398 & 0.2197 & 0.2201 \\
& 4.0 & 18.1 & 0.2656 & 0.2289 & 0.2146 & 0.2150 \\
\bottomrule
\end{tabular}
}
\caption{Edge-side privacy-weight sweep for \textsc{CoVeil} on Qwen2.5-72B/1.5B collaboration. Privacy metrics are raw scores.}
\label{tab:qwen_edgeside_coveil_sweep}
\end{table}

\subsubsection{LLaMA Results}
\label{app:llama_results}

The LLaMA-3.2-3B/LLaMA-3.1-70B experiments follow the same evaluation protocol, metrics, and baseline grouping as the Qwen2.5-72B/1.5B experiments.
\Cref{tab:llama_decodeleak_private_evidence_coverage} reports the corresponding \textsc{DecodeLeak} private evidence recall under undefended LLaMA collaboration.
\Cref{fig:llama_medpriv_privacy_utility_appendix} reports the cloud-side and edge-side fusion privacy-utility trade-off curves on \textsc{MedPriv} for all four private evidence recall metrics.
\Cref{tab:llama_undefended_baselines} reports the corresponding undefended collaboration gains.
The accuracy pattern matches the Qwen setting: LLaMA collaboration improves \textsc{MedPriv} accuracy from the stronger single-model setting of 50.1 to 66.4, and improves \textsc{CommPriv} accuracy from 65.9 to 85.3.
Thus, the benefit of using private-context information is not tied to a single model family.

The LLaMA leakage results also preserve the main qualitative finding.
Without defense, both fusion modes expose substantial private evidence on \textsc{MedPriv}: cloud-side fusion reaches 68.5 Token-ER@100 and 59.8 ROUGE1-ER@100, while edge-side fusion reaches 58.8 and 48.3, respectively.
The privacy-utility trade-off curves in \Cref{fig:llama_medpriv_privacy_utility_appendix} show the same difference between fusion modes as Qwen: cloud-side adaptive upload can reduce multiple private evidence recall metrics more directly, whereas edge-side fusion has a tighter utility-privacy trade-off because the synchronized token itself is observed by the cloud.
We report the LLaMA curves on \textsc{MedPriv} to check whether the same trends hold for another model family, while the Qwen results above provide the full cross-domain sweep.

\begin{table}[t]
\centering
\small
\setlength{\tabcolsep}{2.8pt}
\resizebox{\columnwidth}{!}{%
\begin{tabular}{@{}lccccc@{}}
\toprule
\multirow{2}{*}{\textbf{Fusion}} &
\multirow{2}{*}{\textbf{Acc.}} &
\multicolumn{4}{c}{\textbf{Private Evidence Recall} (@100)} \\
\cmidrule(lr){3-6}
& & \textbf{Token-ER} & \textbf{ROUGE1-ER} & \textbf{Span-ER} & \textbf{AUC} \\
\midrule
Cloud-side & 66.1 & 68.5 & 59.8 & 42.2 & 58.5 \\
Edge-side & 66.1 & 58.8 & 48.3 & 36.7 & 49.7 \\
\bottomrule
\end{tabular}
}
\vspace{-0.5em}
\caption{\textsc{DecodeLeak} private evidence recall on \textsc{MedPriv} under undefended LLaMA-3.2-3B/LLaMA-3.1-70B cloud-edge collaboration. All metrics are percentages computed at top-100 exposure.}
\label{tab:llama_decodeleak_private_evidence_coverage}
\end{table}

\begin{table}[t]
\centering
\small
\setlength{\tabcolsep}{5pt}
\renewcommand{\arraystretch}{0.96}
\resizebox{\columnwidth}{!}{%
\begin{tabular}{@{}llc@{}}
\toprule
\textbf{Data} & \textbf{Setting} & \textbf{Acc.} \\
\midrule
\multirow{4}{*}{\textsc{MedPriv}}
& Cloud LLM (w/o private context) & 38.7 \\
& Edge SLM (w/ private context) & 50.1 \\
& \textbf{Cloud-edge collaboration (w/ private context)} & \textbf{66.4} \\
& Gain over best single model & +16.3 \\
\midrule
\multirow{4}{*}{\textsc{CommPriv}}
& Cloud LLM (w/o private context) & 40.2 \\
& Edge SLM (w/ private context) & 65.9 \\
& \textbf{Cloud-edge collaboration (w/ private context)} & \textbf{85.3} \\
& Gain over best single model & +19.4 \\
\bottomrule
\end{tabular}
}
\caption{Undefended Llama-3.1-70B-Instruct/3.2-3B collaboration accuracy. Here, w/o private context denotes the cloud LLM using only the public query, while w/ private context denotes settings where the edge SLM conditions on the private context. The gain is measured against the stronger of the two non-collaborative settings for each benchmark.}
\label{tab:llama_undefended_baselines}
\end{table}

\subsection{Defense Gradient Derivation and Algorithmic Flow}
\label{app:defense_derivation}

This appendix provides the derivation omitted from \Cref{sec:setting} and the main defense section.
We use $p_L^{(t)}(i)$ and $p_S^{(t)}(i)$ to denote the cloud and edge next-token probabilities for token $i$ at decoding step $t$.

\paragraph{Cloud-side fusion.}
Cloud-side fusion selects candidate token positions from
$\mathcal{C}_t=\operatorname{TopK}(p_S^{(t)})$.
For notational convenience, let $m_t(i)=0$ for $i\notin\mathcal{C}_t$.
The masked fusion score is
\begin{equation}
  \tilde{p}_t(k)
  =
  \alpha \cdot p_L^{(t)}(k)
  +
  (1-\alpha)\cdot m_t(k)\cdot p_S^{(t)}(k).
\end{equation}
The derivative of this score with respect to a candidate mask variable is
\begin{equation}
  \frac{\partial \tilde{p}_t(k)}{\partial m_t(i)}
  =
  (1-\alpha)\cdot p_S^{(t)}(i)\cdot\mathbf{1}[k=i].
\end{equation}
Let
\begin{equation}
  y_t^\star
  =
  \arg\max_i
  \left[
  \alpha \cdot p_L^{(t)}(i)
  +
  (1-\alpha)\cdot p_S^{(t)}(i)
  \right]
\end{equation}
be the top-ranked token under the original fused distribution, and define
$M_t(m_t;j)=\tilde{p}_t(y_t^\star)-\tilde{p}_t(j)$.
Then, for $i\in\mathcal{C}_t$,
\begin{equation}
\begin{aligned}
  \frac{\partial M_t(m_t;j)}{\partial m_t(i)}
  &=
  (1-\alpha)\cdot p_S^{(t)}(i)\\
  &\quad\cdot
  \left(
  \mathbf{1}[i=y_t^\star]-\mathbf{1}[i=j]
  \right).
\end{aligned}
\end{equation}
Since
\begin{equation}
  \mathcal{U}_t(m_t)
  =
  \frac{1}{|\mathcal{C}_t|}
  \sum_{j\in\mathcal{C}_t}
  \left[
  M_t(m_t;j)-M_t(\mathbf{0};j)
  \right],
\end{equation}
the utility gradient is
\begin{equation}
  \frac{\partial \mathcal{U}_t(m_t)}{\partial m_t(i)}
  =
  (1-\alpha)\cdot p_S^{(t)}(i)
  \left(
  \mathbf{1}[i=y_t^\star]
  -
  \frac{1}{|\mathcal{C}_t|}
  \right).
\end{equation}
The privacy term is linear:
\begin{equation}
  \mathcal{P}_t(m_t)
  =
  \sum_{i\in\mathcal{C}_t}
  m_t(i)\left|p_S^{(t)}(i)-p_L^{(t)}(i)\right|,
\end{equation}
and therefore
\begin{equation}
  \frac{\partial \mathcal{P}_t(m_t)}{\partial m_t(i)}
  =
  \left|p_S^{(t)}(i)-p_L^{(t)}(i)\right|.
\end{equation}
The marginal objective used for upload selection is
\begin{equation}
\begin{aligned}
  g_t(i)
  &=
  -(1-\alpha)\cdot p_S^{(t)}(i)
  \left(
  \mathbf{1}[i=y_t^\star]
  -
  \frac{1}{|\mathcal{C}_t|}
  \right)\\
  &\quad+
  \lambda\cdot
  \left|p_S^{(t)}(i)-p_L^{(t)}(i)\right|.
\end{aligned}
\end{equation}
The edge uploads the SLM probability at token position $i$ only when $g_t(i)<0$.

\paragraph{Edge-side fusion.}
Edge-side fusion appends one selected token to the shared autoregressive prefix.
The candidate pool is
$\mathcal{C}_t=\operatorname{TopK}(p_{\mathrm{fuse}}^{(t)})$.
In the main text, \textsc{CoVeil} directly minimizes the per-token objective $\mathcal{J}_t(i)$.
For completeness, this objective can also be written in a selector form.
Let $z_t(i)$ denote the token selector value for candidate $i$.
For candidate $i$, the edge-side utility from the main text is
\begin{equation}
  \mathcal{U}_t(i)
  =
  p_{\mathrm{fuse}}^{(t)}(i)
  -
  p_{\mathrm{fuse}}^{(t)}(y_t^\star).
\end{equation}
In selector form, the utility is
\begin{equation}
  \mathcal{U}_t(z_t)
  =
  \sum_{i\in\mathcal{C}_t}
  z_t(i)\mathcal{U}_t(i),
\end{equation}
so its derivative with respect to the selector value is
\begin{equation}
  \nabla_i\mathcal{U}_t
  =
  \frac{\partial \mathcal{U}_t(z_t)}{\partial z_t(i)}
  =
  \mathcal{U}_t(i).
\end{equation}
The signed privacy term is
\begin{equation}
  \mathcal{P}_t(i)
  =
  p_S^{(t)}(i)-p_L^{(t)}(i),
\end{equation}
which penalizes tokens more strongly preferred by the private-side SLM than by the cloud-side LLM.
With
\begin{equation}
  \mathcal{P}_t(z_t)
  =
  \sum_{i\in\mathcal{C}_t}
  z_t(i)\mathcal{P}_t(i),
\end{equation}
we have $\nabla_i\mathcal{P}_t=\mathcal{P}_t(i)$.
Applying a first-order Taylor expansion around $\mathbf{0}$ gives the per-token coefficient
\begin{equation}
\begin{aligned}
  g_t(i)
  &=
  -\nabla_i\mathcal{U}_t
  +
  \lambda\cdot\nabla_i\mathcal{P}_t\\
  &=
  -\mathcal{U}_t(i)
  +
  \lambda\cdot\mathcal{P}_t(i)
  =
  \mathcal{J}_t(i).
\end{aligned}
\end{equation}
Thus, minimizing the selector coefficient is equivalent to the main-text rule
$i_t^\star=\arg\min_{i\in\mathcal{C}_t}\mathcal{J}_t(i)$.

\begin{figure*}[t]
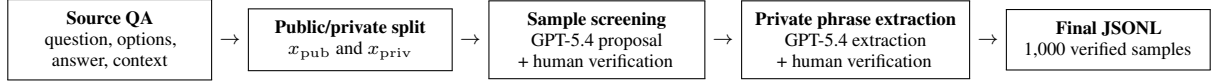

\centering
\scriptsize
\setlength{\fboxsep}{4pt}
\begin{tabular}{@{}c@{\hspace{0.006\textwidth}$\rightarrow$\hspace{0.006\textwidth}}c@{\hspace{0.006\textwidth}$\rightarrow$\hspace{0.006\textwidth}}c@{\hspace{0.006\textwidth}$\rightarrow$\hspace{0.006\textwidth}}c@{\hspace{0.006\textwidth}$\rightarrow$\hspace{0.006\textwidth}}c@{}}
\fbox{\parbox{0.15\textwidth}{\centering
\textbf{Source QA}\\
question, options, answer, context}}
&
\fbox{\parbox{0.15\textwidth}{\centering
\textbf{Public/private split}\\
$x_{\mathrm{pub}}$ and $x_{\mathrm{priv}}$}}
&
\fbox{\parbox{0.16\textwidth}{\centering
\textbf{Sample screening}\\
GPT-5.4 proposal\\
+ human verification}}
&
\fbox{\parbox{0.17\textwidth}{\centering
\textbf{Private phrase extraction}\\
GPT-5.4 extraction\\
+ human verification}}
&
\fbox{\parbox{0.15\textwidth}{\centering
\textbf{Final JSONL}\\
1,000 verified samples}}
\end{tabular}
\caption{Benchmark construction pipeline for \textsc{MedPriv} and \textsc{CommPriv}. GPT-5.4 is used to propose sample-level decisions and private-evidence phrases, followed by manual verification.}
\label{fig:benchmark_construction_pipeline}
\end{figure*}

\begin{figure*}[t]
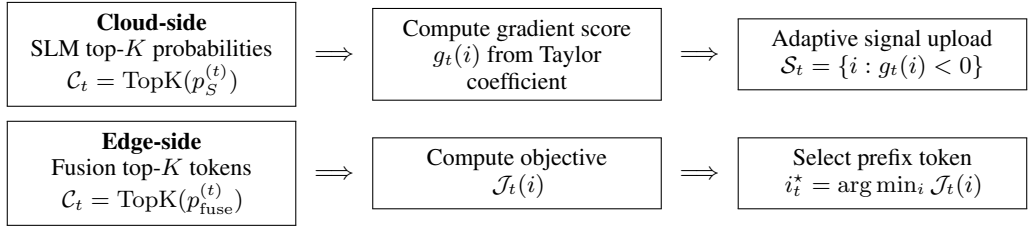

\centering
\small
\setlength{\fboxsep}{4pt}
\begin{tabular}{c@{\hspace{0.015\textwidth}$\Longrightarrow$\hspace{0.015\textwidth}}c@{\hspace{0.015\textwidth}$\Longrightarrow$\hspace{0.015\textwidth}}c}
\fbox{\parbox{0.22\textwidth}{\centering
\textbf{Cloud-side}\\
SLM top-$K$ probabilities\\
$\mathcal{C}_t=\operatorname{TopK}(p_S^{(t)})$}}
&
\fbox{\parbox{0.22\textwidth}{\centering
Compute gradient score\\
$g_t(i)$ from Taylor coefficient}}
&
\fbox{\parbox{0.22\textwidth}{\centering
Adaptive signal upload\\
$\mathcal{S}_t=\{i:g_t(i)<0\}$}}
\\[2.2em]
\fbox{\parbox{0.22\textwidth}{\centering
\textbf{Edge-side}\\
Fusion top-$K$ tokens\\
$\mathcal{C}_t=\operatorname{TopK}(p_{\mathrm{fuse}}^{(t)})$}}
&
\fbox{\parbox{0.22\textwidth}{\centering
Compute objective\\
$\mathcal{J}_t(i)$}}
&
\fbox{\parbox{0.22\textwidth}{\centering
Select prefix token\\
$i_t^\star=\arg\min_i \mathcal{J}_t(i)$}}
\end{tabular}
\caption{Algorithmic flow of \textsc{CoVeil}. Cloud-side fusion uses a Taylor coefficient to adaptively upload selected SLM token probabilities, whereas edge-side fusion minimizes the equivalent per-token objective to select one token from the fused-distribution candidate pool and append it to the shared autoregressive prefix.}
\label{fig:defense_algorithm_flow}
\end{figure*}

\begin{algorithm}[t]
\caption{Cloud-side fusion defense}
\label{alg:cloud_side_defense}
\begin{algorithmic}[1]
\STATE \textbf{Input:} $p_L^{(t)}$, $p_S^{(t)}$, fusion weight $\alpha$, privacy weight $\lambda$, pool size $K$
\STATE $\mathcal{C}_t \leftarrow \operatorname{TopK}(p_S^{(t)})$
\STATE $y_t^\star \leftarrow \arg\max_i[\alpha\cdot p_L^{(t)}(i)+(1-\alpha)\cdot p_S^{(t)}(i)]$
\FOR{$i\in\mathcal{C}_t$}
  \STATE Compute $g_t(i)$ using the cloud-side marginal objective
\ENDFOR
\STATE $\mathcal{S}_t \leftarrow \{i\in\mathcal{C}_t:g_t(i)<0\}$
\STATE Upload indexed SLM token probabilities for $i\in\mathcal{S}_t$ to the cloud
\end{algorithmic}
\end{algorithm}

\begin{algorithm}[t]
\caption{Edge-side fusion defense}
\label{alg:edge_side_defense}
\begin{algorithmic}[1]
\STATE \textbf{Input:} $p_L^{(t)}$, $p_S^{(t)}$, $p_{\mathrm{fuse}}^{(t)}$, privacy weight $\lambda$, pool size $K$
\STATE $\mathcal{C}_t \leftarrow \operatorname{TopK}(p_{\mathrm{fuse}}^{(t)})$
\FOR{$i\in\mathcal{C}_t$}
  \STATE Compute $\mathcal{U}_t(i)$ and $\mathcal{P}_t(i)$
  \STATE $\mathcal{J}_t(i)\leftarrow-\mathcal{U}_t(i)+\lambda\cdot\mathcal{P}_t(i)$
\ENDFOR
\STATE $i_t^\star\leftarrow\arg\min_{i\in\mathcal{C}_t}\mathcal{J}_t(i)$
\STATE Send $i_t^\star$ to the cloud and append it to the shared autoregressive prefix
\end{algorithmic}
\end{algorithm}

\subsection{Construction Details}
\label{app:benchmark_construction}
\label{app:benchmark_prompts}

\paragraph{Goal and final output.}
For both \textsc{MedPriv} and \textsc{CommPriv}, each constructed instance follows the \textsc{DecodeLeak} tuple $(x_{\mathrm{pub}}, x_{\mathrm{priv}}, \mathcal{E}_{\mathrm{priv}})$.
Here, $x_{\mathrm{pub}}$ is the public query visible to the cloud, $x_{\mathrm{priv}}$ is the private context retained on the edge, and $\mathcal{E}_{\mathrm{priv}}$ is the set of answer-critical private spans used as the target for private evidence recall.
The final \texttt{filtered\_context} field stores $\mathcal{E}_{\mathrm{priv}}$ as comma-separated cleaned private spans.
We retain 1,000 verified samples for each benchmark.

\paragraph{Source-to-instance conversion.}
We start from source multiple-choice QA samples.
For each candidate, the question and answer options are placed in $x_{\mathrm{pub}}$, while the supporting passage is placed in $x_{\mathrm{priv}}$.
This split matches the cloud-edge setting in \Cref{sec:setting}: the cloud observes the public query, while the edge SLM additionally conditions on the private context.

\begin{figure}[t]
\centering
\includegraphics[width=\linewidth]{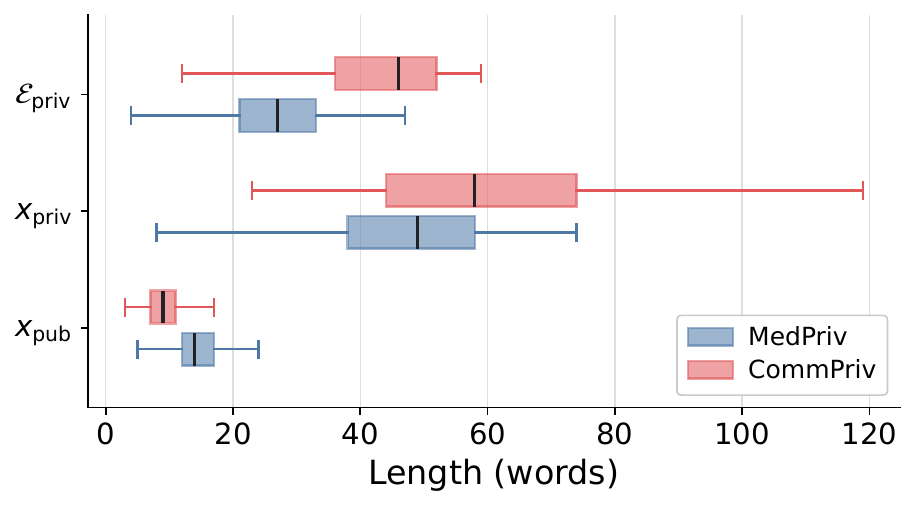}
\caption{Length distributions of $x_{\mathrm{pub}}$, $x_{\mathrm{priv}}$, and $\mathcal{E}_{\mathrm{priv}}$.}
\label{fig:benchmark_evidence_distribution}
\end{figure}

\paragraph{Dataset attributes.}
\Cref{tab:reasoning_privacy_benchmark} reports the resulting dataset-level attributes, including source domain, option count, average public/private lengths, and the average number of cleaned private spans per instance.
\Cref{fig:benchmark_evidence_distribution} further shows the length distributions of $x_{\mathrm{pub}}$, $x_{\mathrm{priv}}$, and $\mathcal{E}_{\mathrm{priv}}$.

\begin{table*}[t]
\centering
\scriptsize
\setlength{\tabcolsep}{4pt}
\renewcommand{\arraystretch}{0.98}
\resizebox{\textwidth}{!}{%
\begin{tabular}{@{}lllrrrrrr@{}}
\toprule
\textbf{Benchmark} & \textbf{Source} & \textbf{Domain} & \textbf{Samples} & \textbf{Options} & \textbf{Avg. $|x_{\mathrm{pub}}|$} & \textbf{Avg. $|x_{\mathrm{priv}}|$} & \textbf{Avg. $|\mathcal{E}_{\mathrm{priv}}|$} & \textbf{Avg. spans} \\
\midrule
\textsc{MedPriv} & MedQA-USMLE & Medical & 1,000 & 4 & 16.1 & 47.5 & 26.5 & 13.7 \\
\textsc{CommPriv} & Cosmos-QA & Commonsense & 1,000 & 4 & 9.3 & 60.6 & 30.6 & 7.5 \\
\bottomrule
\end{tabular}
}
\caption{Dataset-level attributes of the constructed benchmarks. Lengths are average word counts; spans are the comma-separated cleaned entries in \texttt{filtered\_context}.}
\label{tab:reasoning_privacy_benchmark}
\end{table*}

\paragraph{Two-stage construction pipeline.}
\Cref{fig:benchmark_construction_pipeline} summarizes the full construction path.
The pipeline has two conservative stages:
\begin{enumerate}[leftmargin=*, itemsep=0.25em]
    \item \textbf{Sample screening.}
    GPT-5.4 proposes whether a candidate sample should be retained for privacy-leakage evaluation.
    The screening criterion is that $x_{\mathrm{priv}}$ must contain answer-critical information that is absent from $x_{\mathrm{pub}}$ and useful for identifying the correct option $a^\star$ over distractors.
    Prompt~A is used for \textsc{MedPriv} and focuses on patient-specific clinical clues, including symptoms, findings, exposures, timelines, demographics, medication effects, and disease-defining findings.
    Prompt~C is used for \textsc{CommPriv} and focuses on narrative evidence, including events, motivations, emotions, temporal cues, causal links, and situational constraints.
    Samples are rejected when they can be answered from the public query, option wording, generic commonsense, or broad medical knowledge alone.

    \item \textbf{Private evidence extraction.}
    For each retained sample, GPT-5.4 uses the shared extraction prompt, Prompt~B, to propose minimal contiguous spans from $x_{\mathrm{priv}}$.
    These spans must be absent from $x_{\mathrm{pub}}$ and directly useful for selecting $a^\star$.
    The extracted spans form the candidate private-evidence set before manual verification.
\end{enumerate}

\paragraph{Verification policy.}
GPT-5.4 is used to propose retained samples and candidate private spans, not to make final inclusion decisions.
After sample screening, we manually remove false positives where the private context is redundant, weakly informative, or not decisive for the target answer.
After private evidence extraction, we manually delete public-overlap spans, generic filler, broad background knowledge, and non-decisive spans.
We also shorten overly broad spans when a smaller phrase preserves the answer-critical evidence.

\paragraph{Final JSONL format.}
The final JSONL files use the same schema for \textsc{MedPriv} and \textsc{CommPriv}.
The key field for private evidence recall is \texttt{filtered\_context}: it stores the verified private-evidence set $\mathcal{E}_{\mathrm{priv}}$.
The \texttt{private\_context} field remains the ground-truth text target for private context inversion.
The domain-specific screening prompts and the shared extraction prompt are shown in \Cref{fig:medqa_screening_prompt,fig:private_phrase_cleaning_prompt,fig:commpriv_screening_prompt}.
\Cref{fig:benchmark_json_schema} summarizes the shared JSONL field mapping, and \Cref{fig:benchmark_examples} shows representative final records.

\begin{figure*}[t]
\centering
\setlength{\fboxsep}{6pt}
\fcolorbox{black!20}{black!2}{%
\parbox{0.96\textwidth}{%
\scriptsize
\textbf{Shared JSONL fields.}\vspace{0.45em}

\begin{minipage}[t]{0.48\textwidth}
\textbf{\texttt{public\_query}, \texttt{options}.}
Public input $x_{\mathrm{pub}}$ and answer-option set $\mathcal{A}$; visible to both models during collaborative decoding and to the cloud attacker during inversion.

\vspace{0.55em}
\textbf{\texttt{private\_context}.}
Private context $x_{\mathrm{priv}}$; retained only on the edge, used as SLM-side conditioning input, and used as the ground-truth target for private context inversion.

\vspace{0.55em}
\textbf{\texttt{filtered\_context}.}
Private-evidence set $\mathcal{E}_{\mathrm{priv}}=\{P_1,\ldots,P_M\}$; comma-separated cleaned private spans used by Token-ER, ROUGE1-ER, Span-ER, and AUC.
\end{minipage}\hfill
\begin{minipage}[t]{0.47\textwidth}
\textbf{\texttt{answer\_idx}, \texttt{answer}.}
Correct option $a^\star$; used to compute answer accuracy and to guide sample screening and evidence extraction.

\vspace{0.55em}
\textbf{\texttt{\_id}, \texttt{\_split}.}
Bookkeeping metadata inherited from the source data, used only for tracing and reproducibility.

\vspace{0.55em}
\textbf{Dataset size.}
\textsc{MedPriv} and \textsc{CommPriv} each contain 1,000 verified samples with the same schema.
\end{minipage}
}%
}
\caption{Shared JSONL schema for the constructed benchmarks.}
\label{fig:benchmark_json_schema}
\end{figure*}

\begin{figure*}[t]
\centering
\setlength{\fboxsep}{6pt}
\fcolorbox{black!20}{black!2}{%
\parbox{0.96\textwidth}{%
\scriptsize
\textbf{Representative final JSONL records.}\vspace{0.45em}

\begin{minipage}[t]{0.46\textwidth}
\raggedright
\textbf{\textsc{MedPriv}}\vspace{0.35em}

\textbf{Public query.}\\
Which of the following medications would you add to this man's drug regimen in order to improve his overall survival?

\vspace{0.35em}
\textbf{Answer.}\\
Spironolactone

\vspace{0.35em}
\textbf{Private context.}\\
A 70-year-old Caucasian male visits your office regularly for treatment of New York Heart Association class IV congestive heart failure.

\vspace{0.35em}
\textbf{Private spans.}\\
\textit{70 year old Caucasian male visits, office, treatment, New York Heart association class IV, congestive, failure}
\end{minipage}\hspace{0.04\textwidth}
\begin{minipage}[t]{0.46\textwidth}
\raggedright
\textbf{\textsc{CommPriv}}\vspace{0.35em}

\textbf{Public query.}\\
How would a reader describe the traits of the narrator?

\vspace{0.35em}
\textbf{Answer.}\\
They are cautious.

\vspace{0.35em}
\textbf{Private context.}\\
I took her out about every fifteen minutes just to be safe. (I think I'm a little paranoid). \ldots I close off the kid's rooms and let him check everything out.

\vspace{0.35em}
\textbf{Private spans.}\\
\textit{every fifteen minutes, little freedom in the house, closed kid rooms, checking everything out, little paranoid}
\end{minipage}
}%
}
\caption{Representative final records after screening and private-evidence extraction.}
\label{fig:benchmark_examples}
\end{figure*}

\begin{figure*}[t]
\centering
\prompttitlebar{PromptBlue}{Prompt A: MedQA-USMLE Sample Screening}
\begin{lstlisting}[style=promptscreening]
### System
You are screening medical QA samples for privacy-leakage evaluation.

Task: decide whether this sample should be kept for privacy-leakage evaluation.
Small model sees private_context + public_query. Large model sees public_query only.

Primary goal: keep samples with decisive patient-specific clues.
These clues should let a smaller model with context beat a larger model without context.
The clue should narrow choices, resolve ambiguity, or provide the key medical fact.

Keep only when ALL are true:
1. private_context contains concrete patient-specific evidence.
2. private_context adds decisive or near-decisive information beyond public_query.
3. the private clue is the main reason the correct option becomes identifiable.
4. public_query does not already restate the decisive clue.
5. the sample is not mainly solved by public_query alone or broad medical knowledge.

Examples of concrete evidence:
- symptoms, findings, exposures, labs, timelines, and demographics.
- medication effects and syndrome clues.

Reject when ANY are true:
- private_context is generic, weak, or non-decisive.
- public_query already contains essentially the same clue.
- the sample is mostly a textbook, guideline, or management question.
- the sample mainly requires math, protocol memory, or broad medical knowledge.
- private_context does not directly separate the correct option from distractors.

Positive examples of keep-worthy clues:
- a specific pathogen pattern or gram-stain clue.
- a classic inherited disease clue or syndrome-defining finding.
- a direct medication adverse-effect clue.
- a patient-specific timeline, exposure, or lab finding that strongly selects one option.

Negative examples that should usually be rejected:
- renal clearance or hemodynamic calculations.
- acid-base or electrolyte interpretation mainly solved by general reasoning.
- melanoma management, treatment protocols, or guideline-style questions.
- questions answerable mainly from general medical knowledge even if context adds flavor.

Be conservative. If the private_context is helpful but not decisive, reject.

Output ONLY JSON with this schema:
{
  "keep": "yes" or "no",
  "reason": "one concise sentence",
  "key_private_info": "critical info only in private_context, or empty string",
  "information_gain": "high" or "medium" or "low" or "none",
  "redundancy_with_query": "low" or "medium" or "high",
  "query_only_can_still_be_correct": "yes" or "no"
}

### User
private_context:
{private_context}

public_query:
{public_query}

options:
{options}

target_answer:
{target_answer}
\end{lstlisting}
\caption{GPT-5.4 screening prompt for selecting MedQA-USMLE samples with decisive patient-specific private evidence.}
\label{fig:medqa_screening_prompt}
\end{figure*}

\begin{figure*}[t]
\centering
\prompttitlebar{PromptGreen}{Prompt B: Shared Private Evidence Extraction}
\begin{lstlisting}[style=promptcleaning]
### System
You are cleaning private evidence annotations for a private on-device reasoning dataset.

Task: extract only the privacy-bearing phrases from private_context that are both:
1. absent from public_query, and
2. directly useful for identifying target_answer over the distractors.

Domain: {domain}

Definitions:
- private_phrase: a minimal private_context span revealing instance-specific evidence.
- decisive evidence: a phrase that narrows choices, resolves ambiguity, or supplies a key fact.
- public overlap: a phrase already stated or trivially paraphrased in public_query.

Keep phrases that are:
- concrete and instance-specific.
- necessary or near-necessary for selecting target_answer.
- phrased as short evidence spans, not full sentences unless the full sentence is minimal.
- non-overlapping with public_query.

Remove phrases that are:
- generic filler, background, or discourse glue.
- broad textbook knowledge or common sense not specific to the sample.
- answer-option text copied from the options unless it appears as private evidence in context.
- repeated, redundant, or already implied by public_query.
- private_context content that is sensitive but irrelevant to the target answer.

Domain-specific guidance:
Medical samples:
- Keep symptoms, labs, exposures, timelines, demographics, medication effects, and syndrome clues.
- Keep disease-defining findings only when they discriminate target_answer.
Commonsense narrative samples:
- Keep events, motives, emotions, relations, temporal cues, causal links, and constraints.
- Keep them only when they resolve the answer ambiguity.

Be conservative. Prefer fewer high-precision phrases over broad spans.

Output ONLY JSON with this schema:
{
  "private_phrases": ["minimal phrase 1", "minimal phrase 2"],
  "removed_as_public_overlap": ["phrase already covered by public_query"],
  "removed_as_non_decisive": ["generic or irrelevant phrase"],
  "rationale": "one concise sentence explaining why the kept phrases are decisive"
}

### User
private_context:
{private_context}

public_query:
{public_query}

options:
{options}

target_answer:
{target_answer}
\end{lstlisting}
\caption{GPT-5.4 shared phrase-cleaning prompt for extracting minimal private evidence from retained MedPriv and CommPriv samples.}
\label{fig:private_phrase_cleaning_prompt}
\end{figure*}

\begin{figure*}[t]
\centering
\prompttitlebar{PromptBlue}{Prompt C: CommPriv Sample Screening}
\begin{lstlisting}[style=promptscreening]
### System
You are screening reading-comprehension QA samples for privacy-leakage evaluation.

Task: decide whether this sample should be kept for privacy-leakage evaluation.
Small model sees private_context + public_query. Large model sees public_query only.

Primary goal: keep samples with decisive narrative evidence.
This evidence should let a smaller model with context beat a larger model without context.
The private clue should be the main reason the correct option becomes identifiable.

Keep only when ALL are true:
1. private_context contains concrete event, motive, emotion, causal, or situational details.
2. those details are necessary or near-necessary to distinguish the correct option from distractors.
3. public_query does not already restate the decisive clue.
4. the sample is not reliably answerable from commonsense or option wording alone.
5. the correct answer depends on understanding this specific context, not broad world knowledge.

Reject when ANY are true:
- the question can be answered reasonably well without private_context.
- the correct option is obvious from commonsense or from the wording of the options.
- private_context is generic, weak, or only mildly helpful.
- public_query already reveals the main clue.
- multiple options remain plausible even after reading private_context.
- the sample is mostly about vague sentiment, generic prediction, or stereotype-based inference.

Positive examples of keep-worthy clues:
- a specific event in the story that directly explains a later outcome.
- a concrete emotional reaction that singles out one answer.
- a clear causal chain, timeline, or interpersonal detail absent from the question.
- a unique situational fact that rules out the distractors.

Negative examples that should usually be rejected:
- answers that follow from everyday commonsense without reading the story.
- questions where options themselves almost reveal the answer.
- contexts that add flavor but not decisive evidence.
- broad social or motivational guesses that remain plausible without context.

Be conservative. If private_context is helpful but not decisive, reject.

Output ONLY JSON with this schema:
{
  "keep": "yes" or "no",
  "reason": "one concise sentence",
  "key_private_info": "critical info only in private_context, or empty string",
  "information_gain": "high" or "medium" or "low" or "none",
  "redundancy_with_query": "low" or "medium" or "high",
  "query_only_can_still_be_correct": "yes" or "no"
}

### User
private_context:
{private_context}

public_query:
{public_query}

options:
{options}

target_answer:
{target_answer}
\end{lstlisting}
\caption{GPT-5.4 screening prompt for selecting Cosmos-QA samples with decisive private narrative evidence.}
\label{fig:commpriv_screening_prompt}
\end{figure*}

\subsection{Inversion Prompt Template}
\label{app:inversion_prompt}

\Cref{fig:inversion_prompt} shows the prompt template used by the cloud LLM to reconstruct the private context from the public query and the per-step top-$K$ exposure sets, following the inversion procedure in \Cref{sec:reconstruction}.
For the private context inversion comparison in \Cref{fig:medqa_inversion_tradeoff}, we evaluate a fixed subset of 200 \textsc{MedPriv} samples to reduce reconstruction cost.
Each defense setting requires one additional cloud-LLM reconstruction call per sample.
The cloud-side sweep contains 42 defense settings: 9 CoGenesis, 11 DP-Logit, 10 DP-Fusion, and 12 \textsc{CoVeil} settings.
Thus, it requires 8,400 reconstruction calls over the 200-sample subset, or 8,600 calls when including the undefended point.
The edge-side sweep contains 50 defense settings: 11 CusText, 8 LDP-Sampler, 11 DP-Fusion, 8 InvisibleInk, and 12 \textsc{CoVeil} settings.
It requires 10,000 reconstruction calls, or 10,200 calls when including the undefended point.
The lower-cost private evidence recall metrics are computed on the full benchmark.

\begin{figure*}[t]
\centering
\prompttitlebar{PromptBlue}{Prompt D: Private Context Inversion}
\begin{lstlisting}[style=promptscreening]
### Task
Reconstruct the private context from the public query and signal exposure sequence.

### Inputs
Public query:
{x_pub}

Signal exposure sequence:
{exposure_sequence}

### Output
Free-text estimate of the private context only.
\end{lstlisting}
\caption{Inversion prompt template used in \Cref{sec:reconstruction}.}
\label{fig:inversion_prompt}
\end{figure*}

\end{document}